\documentclass[10pt,journal,compsoc]{IEEEtran}

\usepackage[T1]{fontenc}

\ifCLASSINFOpdf
   \usepackage[pdftex]{graphicx}
\else
\fi
\usepackage{mathtools}
\usepackage[cmintegrals]{newtxmath}
\usepackage{algorithm}
\usepackage{algorithmic}

\usepackage{pifont}
\newcommand{\cmark}{\ding{51}}%
\newcommand{\xmark}{\ding{55}}%

\usepackage{booktabs} 
\usepackage{enumitem}
\usepackage{subcaption}
\usepackage{multirow}
\begin{document}
%
\title{DeepEdge: A Deep Reinforcement Learning based Task Orchestrator for Edge Computing}
%
%
%

\author{Baris Yamansavascilar, Ahmet Cihat Baktir, Cagatay Sonmez, Atay Ozgovde, and Cem Ersoy}

\maketitle

\begin{abstract}

The improvements in the edge computing technology pave the road for diversified applications that demand real-time interaction. However, due to the mobility of the end-users and the dynamic edge environment, it becomes challenging to handle the task offloading with high performance. Moreover, since each application in mobile devices has different characteristics, a task orchestrator must be adaptive and have the ability to learn the dynamics of the environment. For this purpose, we develop a deep reinforcement learning based task orchestrator, DeepEdge, which learns to meet different task requirements without needing human interaction even under the heavily-loaded stochastic network conditions in terms of mobile users and applications. Given the dynamic offloading requests and time-varying communication conditions, we successfully model the problem as a Markov process and then apply the Double Deep Q-Network (DDQN) algorithm to implement DeepEdge. To evaluate the robustness of DeepEdge, we experiment with four different applications including image rendering, infotainment, pervasive health, and augmented reality in the network under various loads. Furthermore, we compare the performance of our agent with the four different task offloading approaches in the literature. Our results show that DeepEdge outperforms its competitors in terms of the percentage of satisfactorily completed tasks.

\end{abstract}

\begin{IEEEkeywords}
Edge Computing, Task Offloading, Deep Learning, Reinforcement Learning, Deep Reinforcement Learning
\end{IEEEkeywords}

%
\IEEEpeerreviewmaketitle

\section{Introduction}
%
%
%
%
\IEEEPARstart{T}{he} number of end-users increases dramatically that causes tremendous amount of data in networks by using applications which have diverse requirements. Even though the cloud computing solutions have been used to cope with those application requirements, it is insufficient considering the high data rates, real-time response requirements, vast amounts of data, and user mobility \cite{mao2017survey}. Thus, edge computing is recently applied in order to manage the diverse demands of networks \cite{wu2018spatial}.

Edge computing is an umbrella term \cite{baktir2017can} that encompasses several similar technologies including Cloudlet \cite{satyanarayanan2009case}, Mobile Edge Computing (MEC) \cite{chen2015efficient}, and Fog Computing \cite{computing2016fog}. The idea behind the edge computing is to provide services to applications in a Local Area Network (LAN) or Metropolitan Area Network (MAN) without routing the offloaded tasks to the Wide Area Network (WAN) which causes high service delays that may affect QoS. The expectation from an edge network is that the user can offload its tasks, which cannot be performed on the device, to the corresponding edge server. As a result, end-users can rapidly access and exploit computational resources, and the core network is relieved.

\begin{figure}[t]
\centering
\includegraphics[scale=0.12]{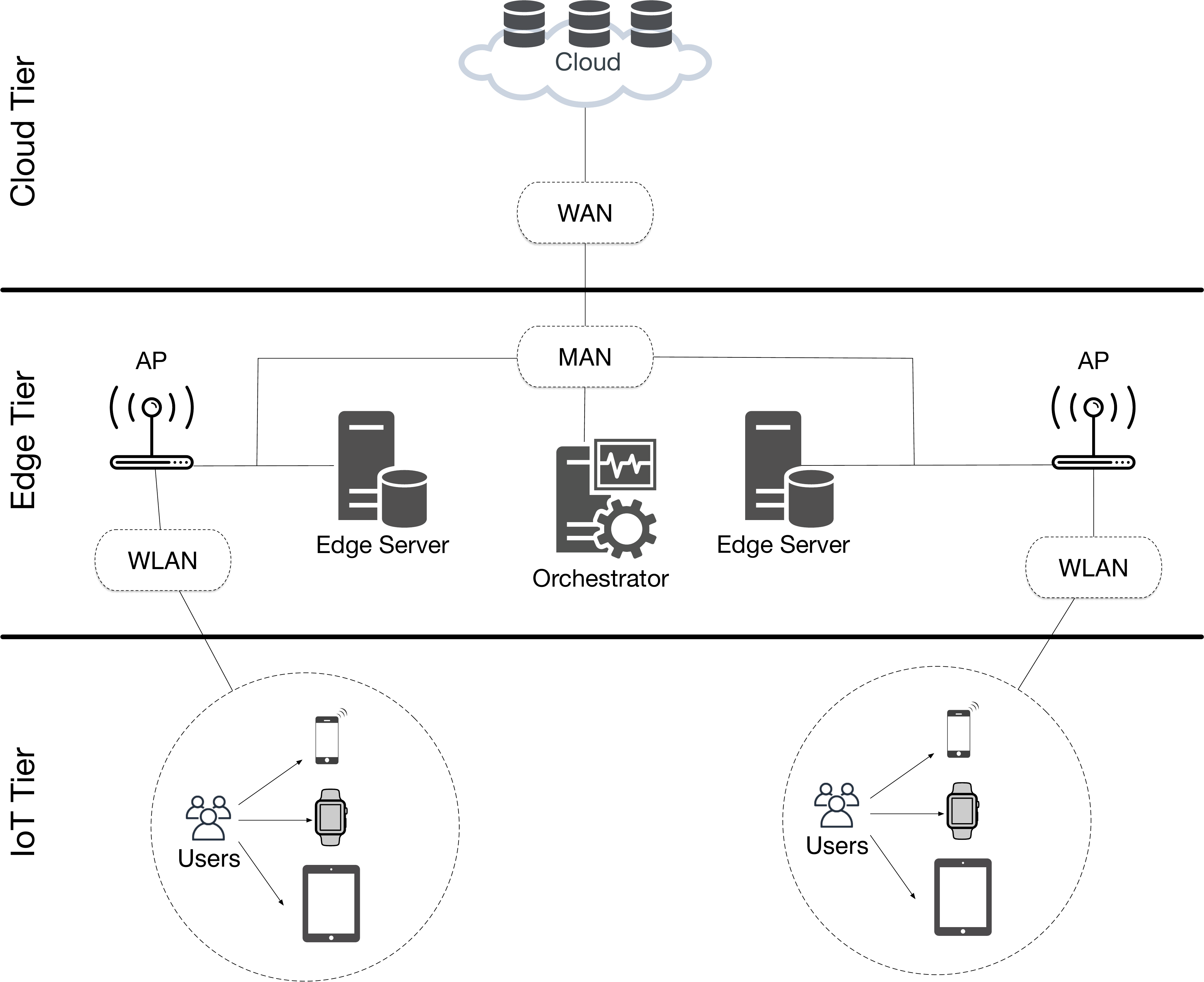}
\caption{A three-tier edge computing network.}
\label{EdgeEnv}
\end{figure}

An edge computing environment is one of the most dynamic and heterogeneous environments since it may consist of multiple edge servers, different services, tasks, technologies, and user profiles. Therefore, deciding where and how the offloaded tasks must be processed in the edge network is crucial. The methodology of this decision is called task orchestration. The primary job of the task orchestrator is to determine whether the offloaded tasks of the users are processed in the cloud via WAN or in the edge via LAN/MAN. Afterwards, if the requested service is available in the edge, the orchestrator should choose the corresponding edge server based on several metrics including delay, length of the task, bandwidth, and utilization of edge servers. A typical architecture of an edge computing environment that consists of the Internet of Things (IoT), edge, and cloud tiers is shown in Figure \ref{EdgeEnv}.

A task orchestrator should be scalable and adaptive since the requirements of the applications and mobile networks may change over time due to their dynamic and heterogeneous nature. Since many parameters must be taken into account, such as delay requirements, length of the task, bandwidth, and utilization of edge servers, traditional rule-based orchestrators would be insufficient to make an accurate decision considering the constraints of the problem. Therefore, a policy must be created in an intelligent way and applied such that the unique demands of the applications are met.

Considering the dynamics of the edge computing networks, the policy for task orchestration should be learned from the environment in order to make accurate decisions. Deep Reinforcement Learning (DRL) is widely used to implement decision-making systems modeled by a Markov Decision Process (MDP) \cite{luong2019applications}. The most important advantage of the DRL is that the agent learns the environment dynamics on its own using the reward mechanism of the system without any human interaction. Therefore, ranging from the communication needs of unmanned aerial vehicles \cite{lin2020dynamic} to  spectrum allocation studies \cite{lei2020deep, li2020deep}, DRL solutions are applied recently in computer networks. Task offloading in edge computing is one of the areas in communication that DRL is also performed \cite{gong2020deep, ning2019deep}. We give their details and also explain what features make our study unique in Section II.

In this study, we introduce our DRL-based task orchestrator, DeepEdge, that is able to meet the dynamic needs of different application types including image rendering, infotainment, pervasive health, and augmented reality under various loads. We designed detailed simulation experiments to evaluate the performance of our proposal in a realistic setting where computational and network level factors are varied. We compared the results with several decision-making systems including the recently developed Fuzzy Logic approach \cite{sonmez2019fuzzy}. Our contributions in this study can be summarized as follows:

\begin{enumerate}

\item DeepEdge can handle the diverse needs of different applications by offloading their tasks into the appropriate edge or cloud server. To the best of our knowledge, the performance evaluation for the application types which have different requirements to meet based on a DRL model has not been performed in the literature.

\item Our DeepEdge orchestrator has learning capability so that it adapts to heavily-loaded environments in terms of mobile users without human interaction. Since studies that apply DRL in the literature do not investigate the effect of the number of mobile users on a decision-making system appropriately, we explore the limits of our DRL agent by deploying varying amount of mobile devices in our simulations.



\item Even though the stochastic nature of the edge computing environment due to different application types and user mobilities cause the delayed action effect, we can model the problem as a Markov process rather than applying policy gradient or semi-Markov process that are used by studies to provide a policy for task offloading. Thereby we can implement the DDQN algorithm for orchestration.

\item We perform online training which is complicated to apply due to stochastic and real-time environment rather than using historical data for DRL.

\end{enumerate}

The rest of this paper is organized as follows. In Section 2, we give a summary of the related works that use deep learning for the task offloading in an edge environment. We formulate the task orchestration problem in Section 3. Section 4 provides the details of DeepEdge including the general architecture along with the training method of the system. In Section 5, we present performance evaluation of our proposed solution. Finally, we conclude our study in Section 6. We list the abbreviations used throughout the paper in Table \ref{listAbbreviations}.

\begin{table}[t]
\caption{List of abbreviations}
\label{listAbbreviations}
\centering
\begin{tabular}{| l | p{6cm} |}
\hline
\textbf{Notation} & \textbf{Description}\\
\hline
DQN &Deep Q-Network \\
DDQN&Double Deep Q-Network\\
DRL&Deep Reinforcement Learning\\
MEC & Mobile Edge Computing \\
LAN & Local Area Network \\
MAN & Metropolitan Area Network \\
WAN & Wide Area Network\\
IoT & Internet of Things\\
MDP&Markov Decision Process\\
QoS& Quality of Service\\
QoE& Quality of Experience\\
VM& Virtual Machine\\
SLA& Service Level Agreement\\
\hline
\end{tabular}
\end{table}

\section{Related Works}

As deep learning solutions provide more accurate results in decision-making systems than the heuristic approaches, they are recently used by studies in edge computing. To this end, the main objective of the Yu et al. \cite{yu2017computation} was to minimize the offloading cost regarding network resources. The authors used a single cell edge computing network and proposed a framework that considers partial computation offloading. They formulated the application, network, and remote/local execution models which were used to generate the training data. They proposed their deep learning model and compared its performance with several offloading schemes in terms of system cost and offloading accuracy. Their scheme was the most successful one among other schemes by reaching up to 60\% accuracy. On the other hand, in \cite{ali2019deep}, Ali et al. built an energy-efficient deep learning based offloading system considering partial tasks. They took the device battery and the energy consumption information into account in order to decide the task was to be offloaded or not. They created a mathematical model based on local and remote execution to generate the training set. They assumed that the decision policies calculated through the mathematical model are 100\% accurate. They compared their approach with several schemes regarding offloading accuracy, energy consumption, and cost based on the number of data samples.

Apart from the traditional DL approaches, different classes in DL are also used by studies in order to provide novel solutions. In \cite{liu2021intelligent}, Liu et al. implemented an intelligent edge-chain enabled access control framework with vehicle nodes, and RoadSide Units (RSUs). In their scenario, the vehicles were lightweight nodes and RSUs were used as edge nodes. The main goal was to perform a secure access control framework for IoV devices. Since such a system requires a historical dataset, they implemented a version of Generative Adversarial Networks (GANs) namely WCGAN. The results showed that their proposed system outperformed the competitors considering the prediction of attacks. However, they did not consider different application/task types even though they did not use historical data as in this study.

Even though DL solutions allow high accuracy, they need historical data which is either labeled for supervised learning or unlabeled for clustering. Moreover, since the correct decision may vary due to the dynamics of the edge computing environment, it is not always possible to train a DNN using traditional methods. Thus, DRL approaches gain attention recently in the literature because of their self-learning capability and high accuracy. In \cite{ning2019deep}, the authors applied a DRL based offloading system for vehicular edge computing. They focused on task scheduling and resource allocation by considering a trade-off between the Quality of Experience (QoE) of the users and the profit of the servers. They split their solution into two modules in which the first one is responsible for the task scheduling, and the second one is used for resource allocation. They implemented DRL for the second module in order to relax the problem. They evaluated the execution time of tasks and the average QoE of users up to 10 vehicles. The results showed that their proposed system is more successful than competitors.

In \cite{zeng2019resource}, the authors used DRL to manage resources at the network edge considering mobility-aware data processing service migration. Their main interest was to lower migration and communication costs between Virtual Machines (VM) and users. To this end, they implemented Deep Q-Network (DQN) algorithm in simulation-based experiments consisting of 50 edge servers and 500 users that randomly move. Their DQN agent outperformed its two competitors in terms of the total cost.

Gong et al. \cite{gong2020deep} considered the trade-off between the local execution cost and task offloading cost regarding energy consumption. To this end, they proposed a hybrid offloading model based on DRL that uses complementary processes of active RF and low-power backscatter communications. Their DRL agent was trained such that it learns to perform optimal transmission method and task offloading between two technologies.

Li et al. \cite{li2018deep} proposed a DRL based framework that aims to minimize the total cost of the delay and energy consumption for all devices in the network. According to the considered local computing and offloading models, the goal of their DRL agent was to find the optimal values of the decision vector and the allocated computational resources to complete the task. They applied their agent in a network that has one edge server and evaluated its performance by comparing them with full offload, full local, and the traditional q-learning considering the total cost. They evaluated the agent up to seven users. 

In \cite{zhu2019computation}, Zhu et al. performed a similar study as \cite{li2018deep} for task offloading by proposing a DRL agent that considers the task completion time and energy consumption. They evaluated the performance of their agent with the local execution and random offloading algorithms regarding those two metrics using one edge server. Based on the different number of components for offloading, their approach outperformed other methods.

Meng et al. \cite{meng2019deep} considered the mean slowdown of the tasks in the queue and the energy consumption. They created a single-user and single edge server setting to evaluate their performance along with the other three offloading schemes including all offload, all local, and random.

In \cite{lu2020optimization}, Lu et al. implemented a DRL-based offloading system that solves the partial task offloading problem for edge computing. Their main goal was to reduce the latency, cost, and energy consumption in the edge network. They improved the DQN algorithm for DRL by using a LSTM network. 
They compared the performance of their system with several DRL solutions and heuristic algorithms based on energy usage, network cost, load balance, and latency metrics. The simulation environment included up to 100 applications, 60 edge servers, and a cloud server. However, the application types and their individual performance were not evaluated, and the number of mobile users in the network was not specified.


In \cite{tang2020deep}, Tang et al. considered the task offloading problem for non-divisible and delay-sensitive tasks. Thus, they used a model-free DRL-based distributed algorithm in which each device takes an action in the environment. To improve the performance of their model, they also include the long short-term memory (LSTM), dueling DQN, and DDQN techniques. In \cite{yan2020optimizing}, the objective was to minimize the average task execution delay and the end device energy consumption on each task by applying a DRL-based joint optimization approach for both device-level and edge-level task offloading. Wu et al. \cite{wu2018spatial} on the other hand, implemented two techniques including decentralized moving edge and multi-tier multi-access edge clustering in order to tackle the mobility, high density, sparse connectivity, and heterogeneity challenges in edge computing. Hence, they used fuzzy logic to jointly consider the multiple inherently contradictory metrics and Q-learning to achieve a self-evolving capability. 

In \cite{carpio2019engineering}, Carpio et al. propose a QoS provider mechanism to work in dynamic scenarios by using a model-free DQN. They optimized QoS by identifying and blocking devices that may cause service disruption due to the dynamicity of the edge computing environment. Alfakih et al. \cite{alfakih2020task} considered the resource management problem in the edge server by performing the optimal offloading decision. Therefore, they implemented on policy RL-based SARSA algorithm in order to minimize the system cost, including energy consumption and computing time delay. In \cite{chen2018optimized}, Chen et al. carried out two double DQN-based online strategic computation offloading algorithms in order to maximize the long-term utility performance of the edge computing environment. The offloading decisions were taken based on the task queue state, the energy queue state as well as the channel qualities between a mobile user and BSs. \cite{huang2019deep} used a DQN-based algorithm in order to take task offloading decisions and carry out wireless resource allocations to the time-varying wireless channel conditions. 

\begin{table*}
  \centering
    \caption{The features of highlighted studies that focused DRL for edge computing}
  \label{featuresOfStudies}
  \begin{tabular}{ c | p{3.2cm} | c | c | c | c   }
  \hline
   \textbf{Study} & \textbf{Solution Method} & \textbf{ \shortstack{Max\\ Mobile Device}} & \textbf{ \shortstack{Max\\ Edge Server}} & \textbf{ \shortstack{Different\\ Task/App Types}} & \textbf{\shortstack{Delayed Action \\ Effect}} \\
   \hline
    \cite{tang2020deep} &  Model-free DQL-based distributed algorithm & 150 & 5 & - & \xmark \\
     \hline
      \cite{yan2020optimizing} & Model-free DQN for both device-level and edge-level task offloading optimizations & 400 & 4 & -& \xmark \\
     \hline
     \cite{carpio2019engineering} & Model-free DQN & 15 & 1 & - & \xmark\\
      \hline
     \cite{wu2018spatial} & Fuzzy logic to jointly consider multiple inherently contradictory metrics and Q-learning to achieve a self-evolving capability. & 500 & 1 - 500 & - & \xmark\\
      \hline
     \cite{alfakih2020task} & On policy SARSA & 5 & 1 & - & \xmark\\
     \hline
     \cite{chen2018optimized} & Two DDQN-based online strategic computation offloading algorithms & 6 & 1 & - & \xmark\\
      \hline
     \cite{liu2020distributed} & Counterfactual multi-agent (COMA) policy gradient, which is a class of actor-critic reinforcement learning approach & 60 & 16 & - & \xmark \\
     \hline
     \cite{huang2019deep} & DQN-based DROO Algorithm & 30 & 1 & - & \xmark\\
     \hline
     \cite{cao2020multiagent} & MADDPG algorithm which is a variant of the actor-critic method for multi-agent environments & 40 & 1 BS with 5 channels & - & \xmark\\
      \hline
     This study & Model free DDQN & 2400 & 14 & 4 & \cmark\\

  \end{tabular}

\end{table*}

Some of the recent studies have focused on the actor-critic algorithms considering each user in the network may take its own offloading decision based on the agent runs on the user device. To this end, Liu et al. \cite{liu2020distributed} implemented counterfactual multi-agent (COMA) policy gradient, which is a class of actor-critic reinforcement learning approach to solve the energy-aware task migration problem. Their goal was to minimize the average completion time of tasks under the migration energy budget. On the other hand, Cao et al. \cite{cao2020multiagent} used multi-agent deep deterministic policy gradients (MADDPG) algorithm \cite{lowe2017multi} to solve the coordination of channel access and task offloading in order to achieve efficient computing.

Considering our extensive examination of the literature, to the best of our knowledge, our study is the first study that investigates the limits of a DRL agent considering the individual application performances and dense environments in terms of the number of mobile devices. Table \ref{featuresOfStudies} shows the main differences between our study and the recent studies that applied DRL. Except for this study, no other studies have examined the performance of a DRL agent considering the different application types that have different features including poisson interarrival times, task size, time sensitivity, etc. This is an important aspect since it increases the level of stochastic nature of the problem in which providing the Markov property is complex. However, in this study, we can successfully ensure the Markov property in an edge computing environment including all application and mobility requirements related to the task offloading. Moreover, considering the max mobile device / max edge server ratio in Table \ref{featuresOfStudies}, we force our agent to take decisions in a very dense environment.




\section{System Model and Problem Formulation}

\begin{table}[t]
\caption{List of symbols}
\label{NotationList}
\centering
\begin{tabular}{| l | p{6cm} |}
\hline
\textbf{Symbol} & \textbf{Description}\\
\hline
$N$ &Number of edge servers\\
$M$ &Number of users \\
$Q$ & Number of tasks \\
$t_{man}$ & MAN delay \\
$t_{wan}$ & WAN delay \\
$t_{q}$ & Total delay of offloaded task q \\
$t_{network}$ & Total network delay \\
$t_{service}$ & Service time of edge or cloud server \\
$t_{SLA}$ & Maximum allowed latency based on SLA of task\\
$C_j$ & j-th edge server capacity \\
$w_q$ & The size of task q\\
$x_{ijq} $ & User i offloads task q to the edge server j\\
$y_{iq}$ & User i offloads task q to the cloud\\
$u_{iq}$ & User i generates task q\\
$\alpha_{q}$ & Success condition for offloaded task q\\

\hline
\end{tabular}
\end{table}

In this section, we introduce the task orchestration problem for the three-tier edge computing environment. The symbols used in the formulation are given in Table \ref{NotationList}.

\subsection{System Overview}

We consider the edge computing environment as a set of edge servers represented by $N_s = \{1, 2, ..., N\}$, a set of mobile users denoted by $M_s = \{1, 2, ..., M\} $, and a set of tasks depicted by $Q_s = \{1, 2, ..., Q\}$. 

In our architecture, there are three tiers as shown in Figure \ref{EdgeEnv}. In the IoT tier, mobile devices generate tasks to be offloaded. The edge tier consists of edge servers and the orchestrator. Each mobile device in the first tier is connected via WLAN to the second tier in which there is an edge server. Finally, in the top tier, the cloud server is placed. Users move according to a nomadic mobility model \cite{ribeiro2011survey} in which a user dwells for a random amount of time in the vicinity of its WLAN/Edge server and then moves to another place in the vicinity of another edge server based on the attractiveness level of the places. If one place's attractiveness level is higher, a user spends more time in that location.

When a mobile device takes the offloading decision, the task is first sent to the orchestrator, which is responsible for the management of the offloaded tasks since it is aware of the recent conditions in the network and edge servers. Thus, depending on the network conditions and task requirements, the orchestrator forwards the task either to an edge server or the cloud server. If the task is successfully processed in the edge or the cloud considering the application requirements in terms of the maximum service time, it means that the orchestration is successful. Otherwise, it is considered as a failure. 


\subsection{Problem Formulation}

In this study, our goal is to minimize the failure rate of the offloaded tasks which is the equivalent of maximizing the success rate of the offloaded tasks. Our objective function can be defined as 

\begin{equation}
	\label{objectiveFunction}
	\max z = \sum_{i=1}^{M}\sum_{q=1}^{Q}u_{iq}\alpha_{q},
\end{equation}
where 
\begin{equation}
   \label{alphaval}
    \alpha_{q}=
    \begin{cases}
      0, & \text{if} \quad t_q > t_{SLA}\\
      1, & \text{otherwise}
    \end{cases}
  \end{equation}
subject to
\begin{equation}
\label{Constraint-1}
	y_{iq} + \sum_{j=1}^{N} x_{ijq} = 1,  \quad  \forall i,q 
	\tag{Constraint 1}
\end{equation}

\begin{equation}
\label{Constraint-2}
	\sum_{i=1}^{M}\sum_{q=1}^{Q}u_{iq} w_q x_{ijq} \leq C_j  \quad \forall j
	\tag{Constraint 2}
\end{equation}


where $u_{iq} \in \{0, 1\}$, $y_{iq} \in \{0, 1\}$, $x_{ijq} \in \{0, 1\}$, $w_q$ is the size of the task $q$, $t_q$ is the total delay of the offloaded task $q$, $t_{SLA}$ is the maximum allowed latency based on the SLA of the task . We define $t_q$ as 
\begin{equation}
\label{totalNetDelay}
	 t_q = t_{service} + t_{network}
\end{equation}

where $t_{service}$ is the service time of the corresponding edge or cloud server, and $t_{network}$ is defined as 

\begin{equation}
   \label{netDelay}
    t_{network}=
    \begin{cases}
     2  t_{wan}, & \text{if} \quad y_{iq} = 1  \quad \forall i, q\\
     2  t_{man}, & \text{if} \quad y_{iq} = 0 \quad \forall i, q
    \end{cases}
  \end{equation}

In Equation \ref{objectiveFunction}, we use $\alpha_{q}$ as an indicator which shows whether the offloaded task is successfully completed or not. If SLA requirements are met based on the offloading decision, $\alpha_{q}$ equals to one (i.e., offloading is successful for that task) as shown in Equation \ref{alphaval}. Otherwise, it is zero as the task fails.

\ref{Constraint-1} represents that a single task for each user can be offloaded to either the cloud server or one of the edge servers. Moreover, \ref{Constraint-2} guarantees that the load of offloaded tasks cannot exceed the capacity of the corresponding edge server.

While calculating the MAN and WAN delays, we use the $M/M/1$ queue model which takes the current network conditions and the requirements of the offloaded task/app type into consideration. If needed, other utilization dependent models can easily be incorporated in the delay models.


\section{DeepEdge}


Mobile networks are one of the most dynamic and heterogeneous environments since they consist of multiple edge servers, different services, tasks, technologies, and user profiles. Therefore, deciding where user tasks are offloaded to is a critical and NP-hard problem \cite{chen2015efficient}. In our design, we assume that a mobile device has already taken the offloading decision. The main goal of our design is to minimize the failed task rate in the network by using the autonomous policy of DRL, even in the heavily-loaded scenarios.

\subsection{Challenges of Applying DRL on Edge}

Even though task orchestration requires a corresponding policy suitable for the concept of DRL, ensuring essential points in a real-time and dynamic environment is a complicated operation. There are four challenges to apply DRL in the edge computing environment:

\begin{itemize}

\item Since tasks are generated by multiple users from multiple applications, guaranteeing the Markov property is a crucial challenge. The transition from one state to its next state should display the change of the environment for the given action that is fundamental for MDP. Therefore, the state representation must be carefully defined in order to apply DRL on edge accurately.

\item The delayed action effect is another imperative challenge due to the fact that the performed action may not immediately affect the network attributes which describe a state for DRL. As a result, there is an important risk that a state and its next state would be the same. An action in MDP must cause a change for its next state like in Atari games \cite{mnih2015human} in order to carry out DRL.

\item The third important challenge is the stochastic environment because of different application tasks. Since task orchestration is modeled considering that each incoming task indicates a state in the system, applying an action would not assure that the next state is the expected state for the edge environment. 

\item The last challenge for this study is that we would like to implement the DRL agent which can learn online rather than using historical data. Related with the delayed action effect, online learning also needs a suitable architecture to implement.

\end{itemize}

\subsection{Providing MDP on Edge }

\begin{figure*}[!t]
\centering
\includegraphics[scale=0.3]{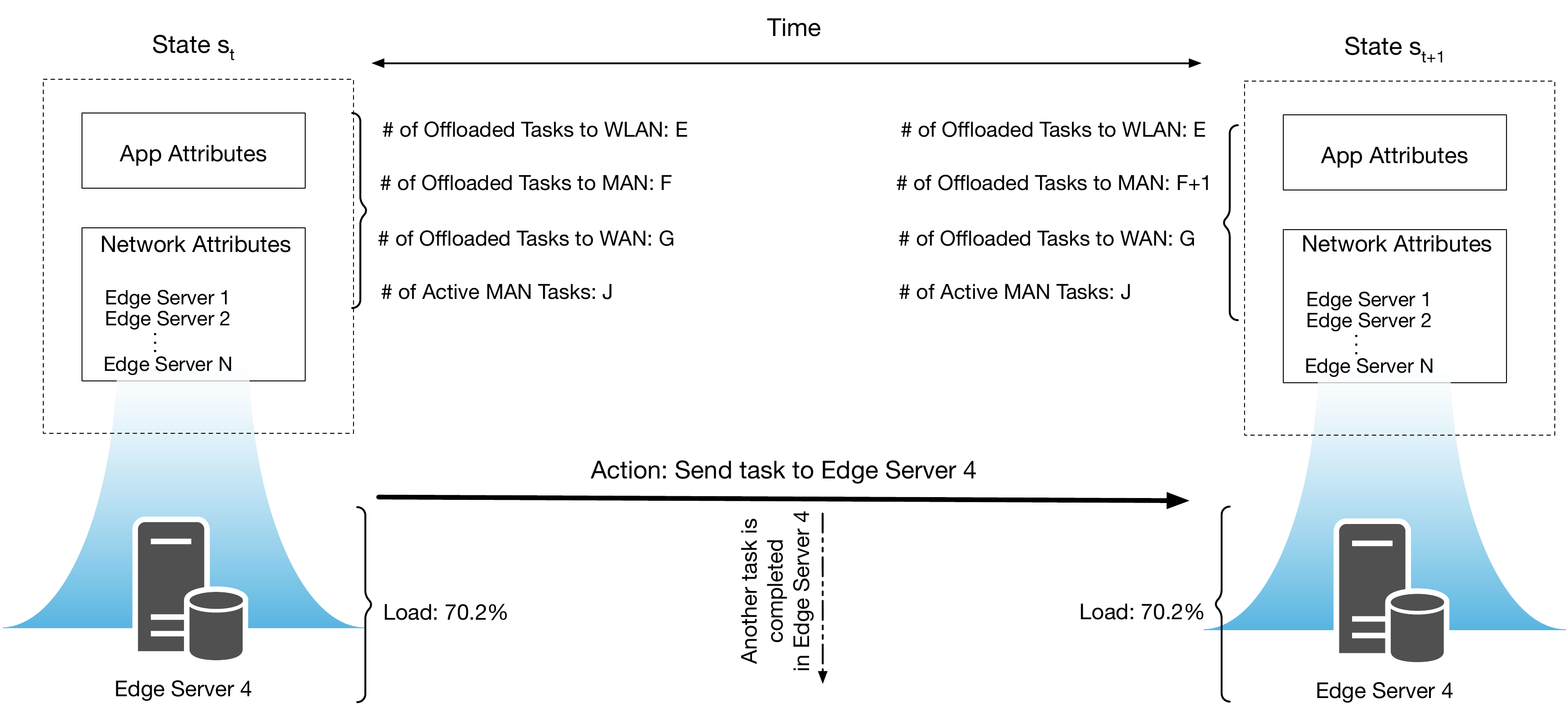}
\caption{Because of the characteristics of the edge computing environment, two consecutive states would be the same regarding the application and network attributes that define the states. This is impractical for DRL since the Markov property may not be ensured. Therefore, we use four supplemental attributes for each state to identify them uniquely.}
\label{mdp-challenge}
\end{figure*}

The DRL is based on the MDP which ensures the mathematical framework to model decision making systems. MDP is formally defined as a 4-tuple $<S, A, P, R>$ where

\begin{itemize}

\item $S$ is the set of all valid states where $s \in S$

\item $A$ is the set of all valid actions where $a \in A$

\item $P: S x A \rightarrow P(S)$ is the state transition probability function that provides the probabilty of $P(s_t, s_{t+1}) = Pr(s_{t+1} = s' \mid s_t = s, a_t = a)$ in which taking the action $a$ as the decision at time $t$ change the current state $s_t$ to $s_{t+1}$.

\item $R: S x A x S \rightarrow R $ is the reward function that defines the reward $r$ for the taken action $a$ at time $t$ considering the state transition such that $r_t = R(s_t, a_t, s_{t+1})$.

\end{itemize}

The agent takes an action based on the observation it gets from the environment so that it can apply a deterministic policy $\pi$ for a given state which is defined as $\pi \left(a \mid s  \right) = P \left(a_t = a \mid s_t  = s \right)$. For the popular implementations of DRL such as Q-Learning, DQN, and DDQN, MDP must be provided for the environment so that the agent can develop a successful policy for its decisions.

MDP is essentially based on the Markov property in which the next state depends only on the current state. In other words, $s_{t+1}$ depends on $s_t$; it is not determined by $s_{t-1}$, $s_{t-2}$, ... , $s_1$, $s_0$. Markov property is formulated as in Equation \ref{mdp}. If \{$s_0$, $s_1$, $s_2$, ...\} is a sequence of discrete random variables, then the sequence is a Markov chain if it satisfies the Markov property. 
\begin{equation}
\label{mdp}
P(s_{t+1} = s \mid s_t, s_{t-1}, s_{t-2}, ..., s_1, s_0) = P(s_{t+1} = s \mid s_t)
\end{equation}

Since the edge computing environment is real-time and extremely dynamic, ensuring Markov property is not a trivial task. As shown in Figure \ref{mdp-challenge}, the orchestrator may send the offloaded task to the Edge Server 4 regarding its load which is 70.2\% at $s_t$. However, if there is another similar task for which the edge server has allocated its resources and it is completed before the arrival of the offloaded task, the load of the edge server would not change at $s_{t+1}$. Moreover, if the $s_{t+1}$ consists of the same application attributes and network attributes, the Markov property cannot be satisfied. Note that this scenario for the load of an edge server is also valid for the load of MAN. This is impractical for DRL that works with the unique states which provide the Markov chain.

In order to solve this problem, we add four supplemental attributes, which are described in Section 4.3 along with other attributes, for each state:

\begin{itemize}

\item Number of Offloaded Tasks to WLAN
\item Number of Offloaded Tasks to MAN
\item Number of Offloaded Tasks to WAN
\item Number of Active MAN Tasks

\end{itemize}


By adding them, we can identify each state uniquely throughout the edge computing environment so that we can satisfy the Markov property for the implementation of DRL.

\subsection{State Representation}

In DRL, a state is defined when the agent takes an action. Considering our system architecture, the agent is the orchestrator that takes action when a task comes to the orchestrator in order to be offloaded to the corresponding edge or the cloud server. Note that since each incoming task is produced by a mobile device by following a Poisson process, the time between consecutive states is exponentially distributed. Thus, we define a state $s$ at which the orchestrator makes a decision for the task.

\begin{figure}[t]
\centering
\includegraphics[scale=0.38]{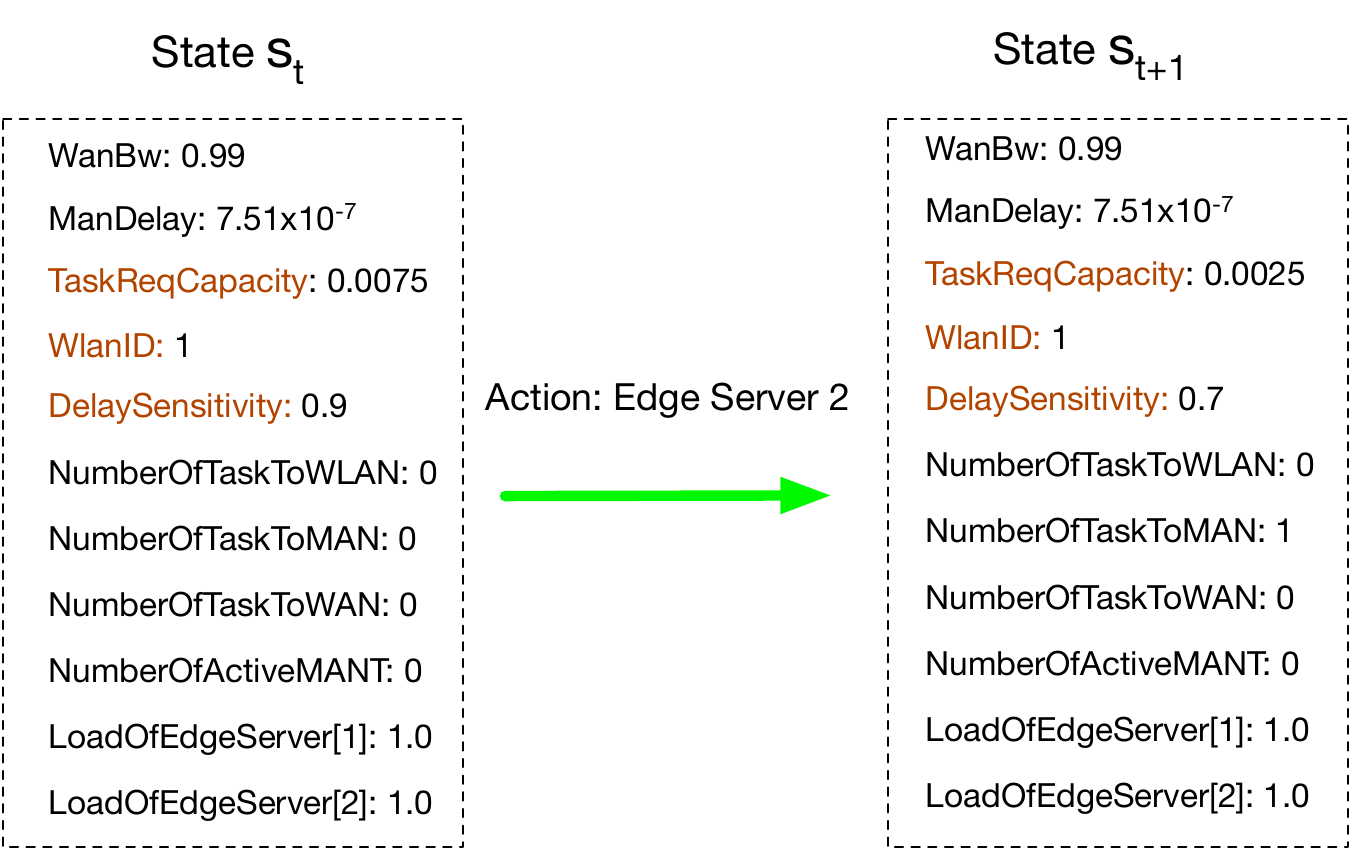}
\caption{The state transition in an edge computing environment consisting of two edge servers.}
\label{StateTransition}
\end{figure}

There are two fundamental feature sets for the state representation in our architecture due to different application requirements: app attributes and network attributes. App attributes indicate the individual requirements of the applications in order to ensure their QoS and to identify the corresponding state. On the other hand, network attributes describe the current condition of the network. Thus, a transition between two consecutive states in a stochastic environment would be as in Figure \ref{StateTransition} and the transition probability $P{ss'}$ is defined as $P{ss'} = P\left(s_{t+1}=s' \mid s_t=s, a_t = a\right)$. However, even though the taken action changes the environment through  four supplemental attributes we add, the action does not fully determine the next state due to the Poisson fashion of the different applications run by multiple users. 

In this study, we use ten features in which three of them indicate app attributes, while seven of them are related to the network condition. On the other hand, since the number of edge servers varies based on the network configuration, the attribute that shows the current load of an edge server depends on the environment. The attributes of a state are given in Table \ref{StateTable}.

For the definition of a state we use \textit{WanBw, ManDelay, NumberOfTaskToWLAN, NumberOfTaskToMAN, NumberOfTaskToWAN, NumberOfActiveMANT}, and \textit{LoadOfEdgeServer[N]} as network attributes. On the other hand, \textit{TaskReqCapacity, DelaySensitivity}, and \textit{WlanID} attributes are used for tasks to indicate their QoS requirements and  location of offloading. \textit{WanBw, ManDelay}, and \textit{LoadOfEdgeServer} refers to the recent conditions of the WAN bandwidth, MAN delay, and load of the edge servers, respectively. Note that,  \textit{LoadOfEdgeServer[N]} includes the information of all of the edge servers in the network so that the $N$ is the variable that represents the number of edge servers. As a result, for example, if there are N edge servers in the network, the corresponding state includes $6 + N$ network related attributes. Hence, along with the application related attributes, the state consists of $9+N$ attributes in total. We demonstrate an example state transition in Figure \ref{StateTransition} considering the scenario in which there are two edge servers in the environment.

The other network attributes including \textit{NumberOfTaskToWLAN, NumberOfTaskToMAN, NumberOfTaskToWAN, NumberOfActiveMANT} are self-explanatory. We increase corresponding values if the tasks are offloaded through WLAN, MAN, and WAN, respectively. Considering application attributes on the other hand, \textit{TaskReqCapacity} indicates the required capacity for the offloaded task regarding edge servers. \textit{DelaySensitivity} refers to the level of application tolerance to delay so that corresponding action should select WLAN edge server, MAN edge server, or the cloud server wisely. Finally, \textit{WlanID} specifies from where the task is offloaded so that the agent can identify which edge server is local for the user.

\begin{table}[!b]
\caption{Attributes of The State}
\label{StateTable}
\centering
\begin{tabular}{|l|p{5cm}|}
\hline
\textbf{Attribute} & \textbf{Description}\\
\hline
\textit{WanBw} & Remaining WAN bandwidth of the network.\\
\hline
\textit{ManDelay} & MAN delay of the network.\\
\hline
\textit{TaskReqCapacity} & Required capacity of a VM to process the given task.\\
\hline
\textit{WlanID} & Wlan ID of mobile device generating the given task.\\
\hline
\textit{DelaySensitivity} & Indicating whether the task is delay intolerant or not. Its value is between 0 and 1.\\
\hline
\textit{NumberOfTaskToWLAN} & The number of tasks offloaded to WLAN. \\
\hline
\textit{NumberOfTaskToMAN} & The number of tasks offloaded to MAN.\\
\hline
\textit{NumberOfTaskToWAN} & The number of tasks offloaded to WAN.\\
\hline
\textit{NumberOfActiveMANT} & The number of tasks offloaded to MAN but have not arrived their destinations yet. \\
\hline
\textit{LoadOfEdgeServer[N]} & The active load of an edge server. There would be N number of edge server in the network.\\

\hline
\end{tabular}
\end{table}

\subsection{Action Space}

The actions of an orchestrator are limited considering the number of edge servers in the network and the cloud server. Therefore, if there are $N$ edge servers, the orchestrator selects one of the $N+1$ decisions as an action $a_t \in \{a_1, a_2, ..., a_{N+1}\}$.

\begin{figure}[t]
\centering
\includegraphics[scale=0.38]{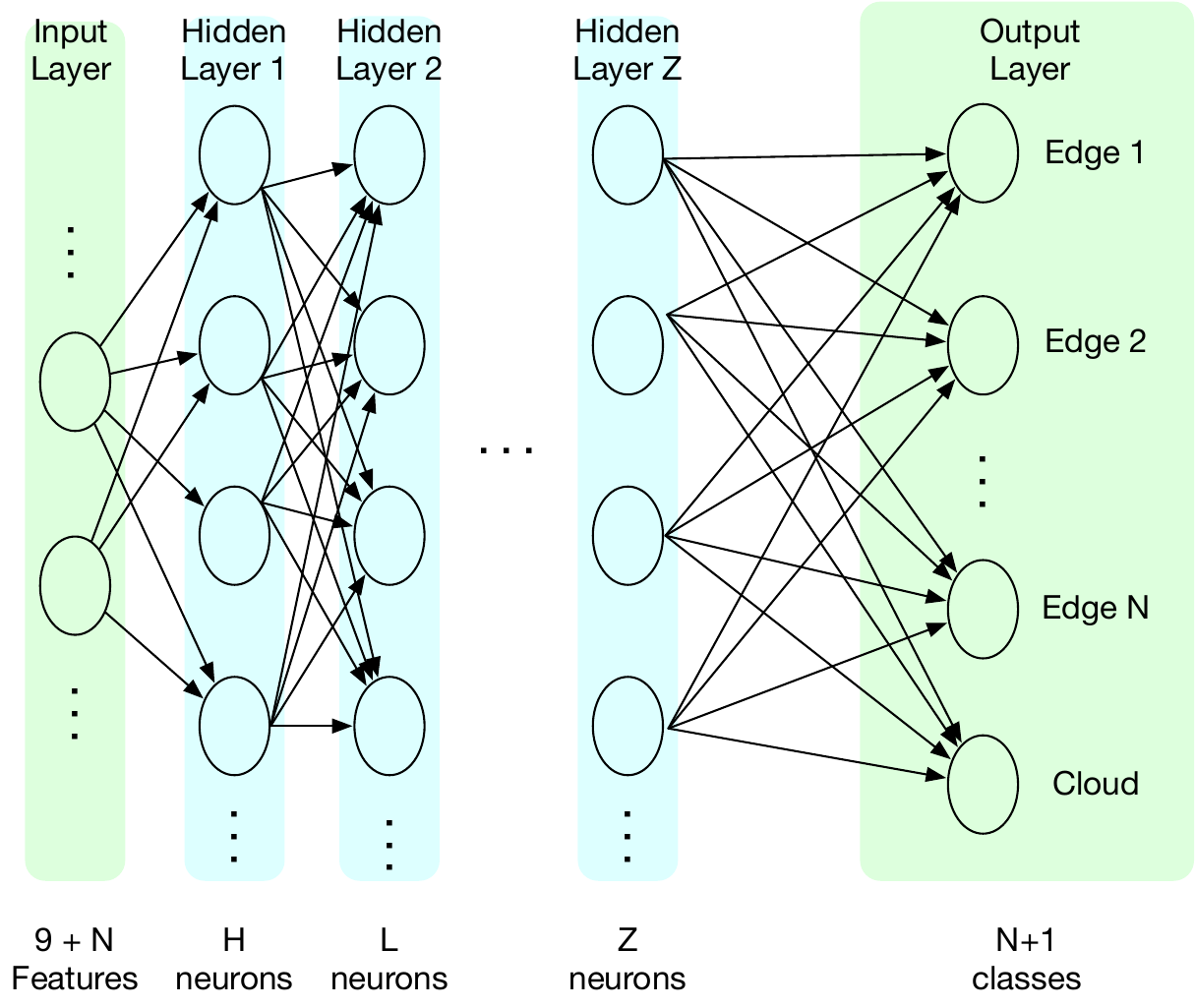}
\caption{The generic deep neural network model for DeepEdge.}
\label{NNGenericForEdge}
\end{figure}

The actions are taken by the neural network part of the orchestrator regarding DRL. When an offloaded task comes to the orchestrator, the current state of the environment is given to the neural network as an input for action. The corresponding generic neural network model is shown in Figure \ref{NNGenericForEdge}. Considering the $N$ edge servers in the network, the input features are defined regarding the state of the environment, which is $9+N$. Based on the network condition and the expected number of users, there would be several hidden layers that consist of multiple neurons. Since the orchestrator determines the corresponding offloading decision among $N+1$ possible choices, as $a_t \in \{a_1, a_2, ..., a_{N+1}\}$, the size of the output layer is equal to the number of edge servers and the cloud server.

\subsection{Reward Mechanism}

As a DRL agent, the long-term goal of the orchestrator is to minimize the failed task rate by making successful offloading decisions. In other words, the orchestrator finds a policy $\pi(s_t)$ that maximizes the expected sum of future rewards at $s_t$ by taking the corresponding action $a_t$ to the $s_{t+1}$, such that

\begin{equation}
\label{ExpectedReward}
R_t =  \sum_{i=t}^{\infty} \gamma*R(s_i, s_{i+1}) 
\end{equation}
where $\gamma \in [0,1]$ is the discount factor to indicate the importance of the immediate or long-term rewards, and $R_t$ is the long-term reward. We define the reward for each state transition so that if the offloaded task is successfully completed, we assign $R(s_t, s_{t+1}) = 1$, otherwise $R(s_t, s_{t+1}) = -1$. 

Our motivation for this reward mechanism is based on our objective function presented in Equation \ref{objectiveFunction}. Since our objective function maximizes the number of successfully offloaded tasks, we formulate the reward function regarding the individual success so that the cumulative reward would reflect the objective function. It is important to note that if our objective function consisted of the minimization of the service time or processing time, the reward mechanism would be different for our system. Moreover, since the orchestrator treats each task equally, it takes the same reward for each offloaded task depending on its decision. We do not apply any prioritization for different applications. However, if needed this could easily be employed.

\subsection{Applying DDQN on Edge }

In order to achieve its objective function of maximizing the expected sum of future rewards, a DRL agent should choose actions under a deterministic policy $\pi(s_t)$ that also maximizes the action-value function, $Q_\pi(s_t , a_t)$, which is also named as the Q-function. Therefore, under a policy $\pi$, the value of an action $a_t$ in a state $s_t$ is

\begin{equation}
\label{QValue}
Q_\pi(s_t , a_t) = \mathop{\mathbb{E}} \big[ R_t \mid s_i = s, a_i = a \big]
\end{equation}
, where $R_t$ is defined in Equation \ref{ExpectedReward}. As a result, an optimal policy is to select the highest valued action in each state such that

\begin{equation}
\label{OptimalPolicy}
\pi(s_t) = arg\max\limits_{a'}(Q(s_t, a'))
\end{equation}
, where $a'$ indicates the set of all possible actions.

In the enhanced DQN algorithm \cite{mnih2015human}, there is a target network along with the experience replay in addition to the online network to carry out a smoother learning process. The target network has the same configuration as the online network in terms of the network size and number of the hidden layers. The only difference between them is the network parameters. The online network parameters, $\theta_t$, are copied to the target network having parameters $\theta^-_t$ for each $\tau$ step, and both of them are fixed in other steps. Thus, the target in DQN is determined as

\begin{equation}
\label{DQN-Value}
z_t = R_{a_t}(s_t, s_{t+1}) + \gamma * Q(s_{t+1}, arg\max\limits_{a'}Q(s_{t+1},a'; \theta^-_t);\theta^-_t)
\end{equation}

However, as investigated in \cite{van2016deep}, the DQN algorithm causes overestimation of the target values since the target network is used for both selecting and evaluating the action. Thus, in this study, we use the DDQN algorithm \cite{van2016deep} in which the online network is used for the selection of the action, and the target network is exploited for evaluating the action as follows

\begin{equation}
\label{DDQN-Value}
z_t = R_{a_t}(s_t, s_{t+1}) + \gamma * Q(s_{t+1}, arg\max\limits_{a'}Q(s_{t+1},a'; \theta_t);\theta^-_t)
\end{equation}

Note that the learning objective of the orchestrator is to minimize the temporal difference error, $\delta_t$, of $Q_\pi(s_t, a_t)$ regarding the target value $z_t$:

\begin{equation}
\label{targerError}
\delta_t = \mid Q(s_t, a_t) - z_t  \mid
\end{equation}

\begin{figure}[!t]
\centering
\includegraphics[scale=0.0755]{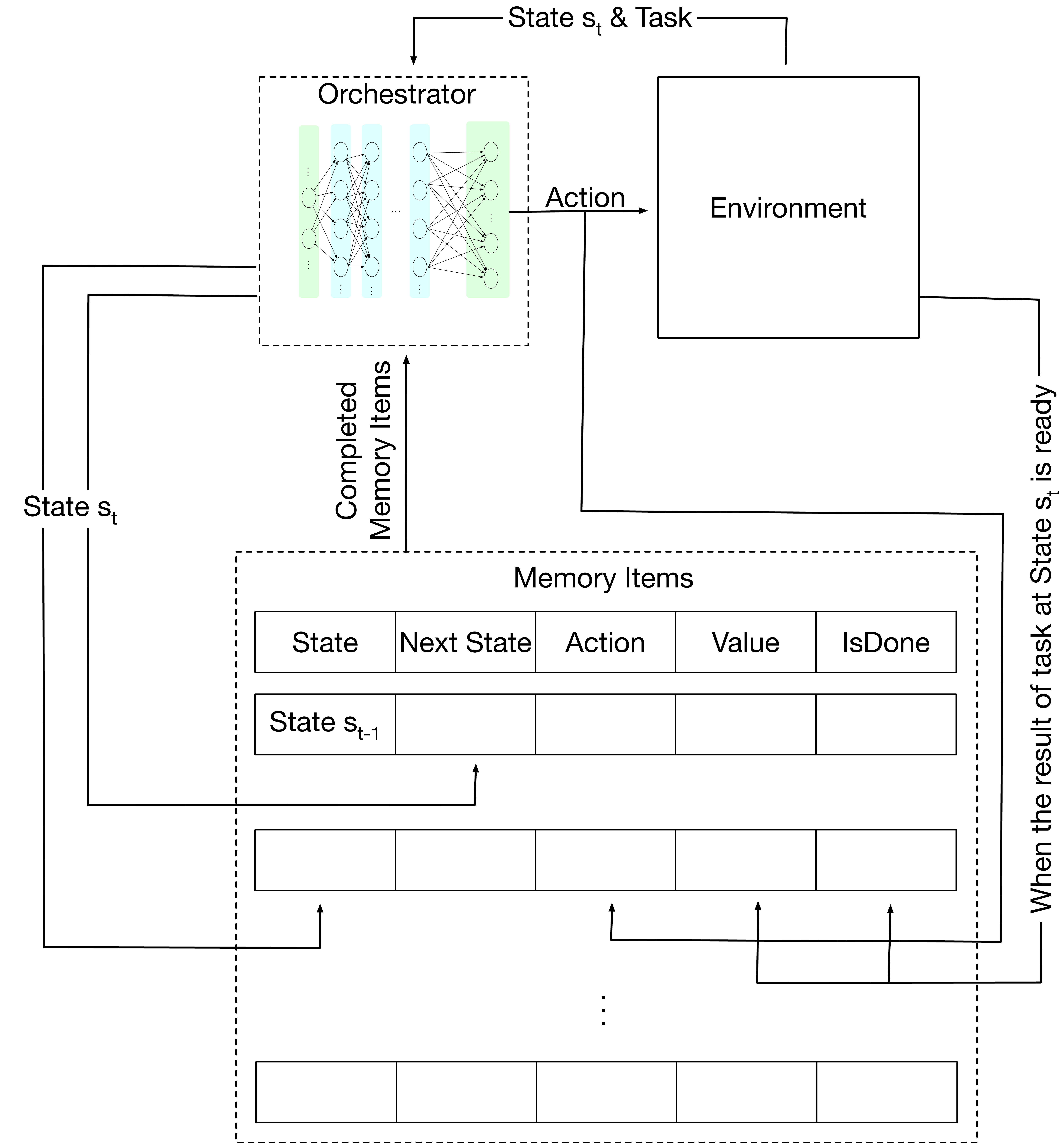}
\caption{5-tuple memory item structure to feed  the DDQN algorithm.}
\label{edgeDQN}
\end{figure}

\subsection{The Delayed Action Effect}

In DRL, the agent gets the next state information from the environment immediately and updates its parameters based on the result of the value function, which is the reward. Since the next state information is a direct result of its action, the agent updates its parameters accurately. However, in our problem, because of the $t_{service}$, $t_{man}$, and $t_{wan}$, the agent cannot immediately get the next state information which is the outcome of its action. This is an important issue since the agent cannot learn the dynamics of the environment as it cannot observe the effect of its actions instantaneously. To solve this delayed action effect, we adapt DDQN into our environment as shown in Figure \ref{edgeDQN}.

In our solution, we use memory items each of which holds 5-tuple information including $state$, $next$ $state$, $action$, $value$, and $isDone$ elements. These elements are essential for the DDQN algorithm in order to feed the agent's memory considering the experience replay and the neural network. After an action is taken, traditionally, they are given to the agent immediately. Since we cannot provide the effect of the action immediately due to our environment, we first insert the $state$ and $action$ elements of the current $task$ into a memory item. Note that since each state is also the next state of its previous state, except for the very first state of the system, we also insert the $state$ into the memory item of the previous task. By using the same fashion, the $next$ $state$ is filled for our current $task$. Afterwards, when the result of the current $task$ is available in the environment, we place the value, regarding the reward mechanism, into the memory item. Moreover, if the task is the last task in the system, we set the $isDone$ element as true. Note that indicating the last element is also imperative in the DDQN algorithm in order to ensure MDP. Finally, if all elements of a memory item are set, it is sent to the memory of the agent considering the experience replay which is used to feed the online network for training. Thus, we can perform online training by overcoming the delayed action effect. We also give the details of our scheme in Algorithm \ref{taskHandlingAlgorithm} and \ref{ApplyingDDQN}, respectively. Since DDQN is a well-known algorithm, we only give the steps of our approach until DDQN is fed by a memory item in Algorithm \ref{ApplyingDDQN}.

In Algorithm \ref{taskHandlingAlgorithm}, our input is the task and our goal, the output, is the creation of the memory item considering the delayed action effect. To this end, we first create an empty memory item in the first line and get the $currentState$. When the agent takes the action, we store this information in the $action$ variable and increase the corresponding values from line 4 to 10 considering to where the task is offloaded. Those values are used as features for the next state. Afterwards, from line 11 to 16, we fill the memory item with the corresponding values. Since we do not know the value and the next state information, $null$ is assigned to those elements initially. Next, we store the $(currentState, memoryItem)$ tuple into the $stateToMemItem$ dictionary in order to change the memory item when we obtain the result of the task. If the current state is not the first state in the environment, we set the next state element of the previous memory item as the $currentState$. Finally, we store the $(task, currentState)$ pair in the $taskToStatePair$ dictionary and then offload the task to the destination based on the selected action.

In Algorithm \ref{ApplyingDDQN}, our input is the completed task and the output is  the completed memory item to feed the agent's memory. Hence, first, the reward is set based on the fail or success condition of the offloaded task. Afterwards, the corresponding state is obtained from the $taskToStatePair$ dictionary in line 6 in order to acquire the memory item related to the state. Next, the $value$, and $isDone$ elements of the memory item is set from line 7 to 12. Finally, if the next state information is set, then the memory item is given to the memory of the agent for the training.

\begin{algorithm} [t]\caption{Handling the incoming tasks considering the delayed action effect} \label{taskHandlingAlgorithm}
\begin{algorithmic}[1]
\REQUIRE $task$ \ENSURE action $a_t$ and memory item of $task$
\STATE $mItem \leftarrow CreateMemoryItem()$
\STATE $currentState \leftarrow GetState()$
\STATE $action \leftarrow agent.doAction(currentState)$ // using online network

\IF{$action = EdgeServerOfTheTask$}
\STATE $NumberOfTaskToWLAN$++
\ELSIF{$action = OtherEdgeServer$}
\STATE $NumberOfTaskToMAN$++
\ELSE
\STATE $NumberOfTaskToWAN$++
\ENDIF
\STATE $mItem.currentState \leftarrow currentState$
\STATE $mItem.nextState \leftarrow null$
\STATE $mItem.value \leftarrow null$
\STATE $mItem.action \leftarrow action$
\STATE $mItem.isDone \leftarrow false$

\STATE $stateToMemItem(currentState.id, mItem)$

\IF{$stateToMemItem(currentState.id-1) \neq null$}
\STATE $prevItem \leftarrow stateToMemItem(currentState.id-1)$
\STATE $prevItem.setNextState(currentState)$
\ENDIF
\STATE $taskToStatePair(task, currentState)$
\STATE $OffloadTaskToDestination(action, task)$
\end{algorithmic}
\end{algorithm}

\begin{algorithm} [t]\caption{Handling completed tasks and memory items}\label{ApplyingDDQN}
\begin{algorithmic}[1]
\REQUIRE A completed $task$ \ENSURE training of neural networks
\IF{$task = isFailed$}
\STATE $reward \leftarrow -1$
\ELSE
\STATE $reward \leftarrow 1$
\ENDIF
\STATE $state \leftarrow taskToStatePair(task)$
\IF{$stateToMemItem(state.id) \neq null$}
\STATE $memItem \leftarrow stateToMemItem(state.id)$
\STATE $memItem.reward \leftarrow reward$
\IF{$task.id = 75000 $} 
\STATE $memItem.isDone \leftarrow true$
\ENDIF

\IF{$memItem.nextState \neq null $} 
\STATE $agent.DDQN(memItem)$ //which save the memory item into the agent's memory for the experience replay
\ENDIF

\ENDIF
\end{algorithmic}
\end{algorithm}

\section{Performance Evaluation}

We carried out our experiments in the EdgeCloudSim \cite{sonmez2017edgecloudsim} simulator for the performance evaluation.
Our scenario consists of 14 locations, each of which operates an edge server with equal capacity. We used different numbers of mobile devices in the edge computing environment to evaluate the performance of the algorithms under variable load. Mobile devices run different applications including augmented reality, pervasive health, image rendering, and infotainment. Each application generates tasks that have different characteristics based on their requirements given in Table \ref{AppChars}. Task interarrival time indicates the average task generation rate. Delay sensitivity refers to an application's tolerance to latency. This value is high for delay-intolerant applications such as video conferencing. VM utilization species the percentage of processing resources required to execute the related task.

\begin{table}[!t]
\caption{Application Properties in the Simulator}
\label{AppChars}
\centering
\begin{tabular}{| p{2.7cm} | p{1.1cm}| p{0.85cm} | p{1cm}| p{0.85cm} |}
\hline
\textbf{Property} & \textbf{\vtop{\hbox{\strut Augment.}\hbox{\strut Reality}}}& \textbf{Perv. Health} & \textbf{\vtop{\hbox{\strut Image}\hbox{\strut Rend.}}}& \textbf{Infot. App}\\
\hline
 Task Interarrival& 2 sec & 3 sec& 20 sec& 7 sec\\
\hline
Delay Sensitivity & 0.9 & 0.7 & 0.1 & 0.3 \\
\hline
Data Upload/Download (KB) & 1500/25 & 20/1250 & 2500/200 & 25/1000 \\
\hline
VM Utilization on Edge (\%)& 6 & 2 & 30& 10 \\
\hline
VM Utilization on Cloud (\%)& 0.6 & 0.2 & 3 & 1 \\
\hline
Usage Percentage (\%)& 30 & 20 & 20 & 30 \\
\hline
\end{tabular}
\end{table}

\subsection{Configuration}

\subsubsection{Simulator Parameters}
We deployed 14 edge servers in the network each of which has 8 VMs, and one cloud server that has 4 VMs. While the computational resource capacity of each VM in an edge server is 10  billion instructions per second (GIPS), a VM in the cloud server has 100 GIPS capacity. For each experiment, the number of mobile devices in the network ranged from 200 to 2400. 
The average number of offloaded tasks is based on both the number of mobile devices and the average task generation rates given in Table \ref{AppChars}. Thus, the expected number of generated tasks would be 6500 and 80000 for 200 and 2400 mobile users, respectively.

We repeated our experiments 40 times with different random seeds. The duration of each experiment in simulator time was five minutes. Since we want to conduct a valid comparison with the competitors, we used the empirical results for the WAN and WLAN delays explained in \cite{sonmez2019fuzzy}. These values can also be observed in EdgeCloudSim. For MAN delay, we used the M/M/1 queueing model implementation of EdgeCloudSim with an additional propagation delay of 5 ms. 


\subsubsection{Training the Agent}

Since we used 14 edge servers in the network, the output layer of the neural networks regarding online and target networks consisted of 15 nodes with respect to our generic model shown in Figure \ref{NNGenericForEdge}. Our final model includes two hidden layers each of which consists of 128 neurons. We applied the ReLU activation function for each neuron in the input and hidden layers. On the other hand, the linear activation function was used for the output layer. We used Stochastic Gradient Descent (SGD) for the optimization of the model. Other important hyperparameters for DDQN are given in Table \ref{Hyperparameters}. In order to define hyperparameters for the $discount$ $factor$, $minibatch$ $size$, and $target$ $network$ $update$ $frequency$, we used random search regarding  our experience in the domain. On the other hand, for the $initial$ $exploration$, $exploration$ $factor$, $final$ $exploration$, and $replay$ $memory$ $size$, we used well-known values from the literature.

Since we use the same agent for the simulations of the different numbers of users, training should provide a robust online neural network considering the different dynamics. Therefore, there are two options: (1) we can train the agent for each network condition regarding 200 to 2400 users, or (2) we can train the agent only in the scenario of 2400 users which theoretically includes the all possible states of the other scenarios. Due to prolonged execution times, we opted for the second option. To give a numeric comparison, completion of a single episode took over three hours for the first option whereas it took around 0.3 hours for the second option. It took 101 episodes (over 30 hours) to train the agent and converge to the range of 10\% - 15\% failed task rate.

\begin{table}[!t]
\caption{Hyperparameters and Their Values}
\label{Hyperparameters}
\centering
\begin{tabular}{|l|l|}
\hline

\textbf{Hyperparameter} & \textbf{Value}\\
\hline
Learning Rate & 0.0001\\
\hline
Discount Factor & 0.8\\
\hline
Initial Exploration & 1\\
\hline
Exploration Factor & 0.99\\
\hline
Final Exploration & 0.1\\
\hline
Replay Memory Size & 1000000\\
\hline
Minibatch Size & 4\\
\hline
Target Network Update Frequency & 10\\

\hline
\end{tabular}
\end{table}

\begin{figure}[t]
\centering
\includegraphics[scale=0.43]{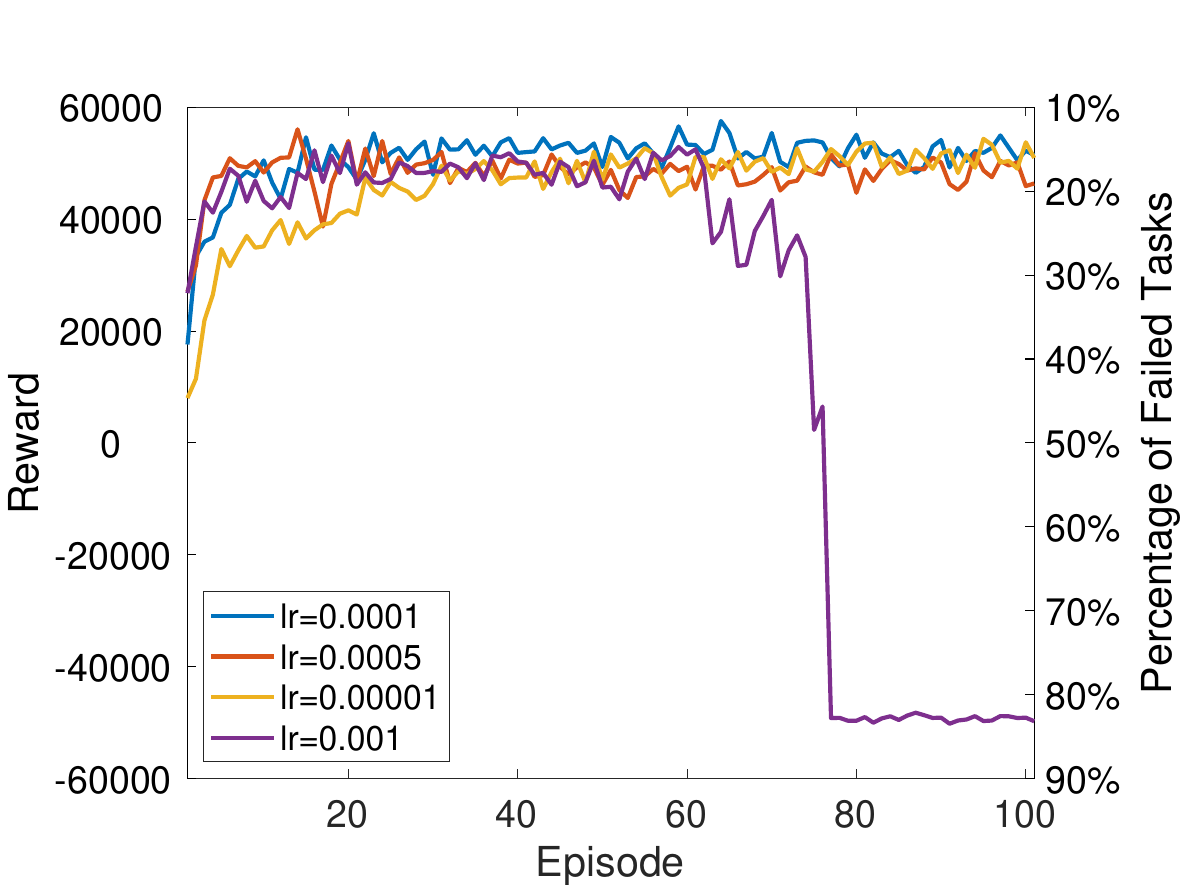}
\caption{Change of the failed task rate over episodes for the scenario of 2400 mobile users.}
\label{scores}
\end{figure}

Figure \ref{scores} shows the rewards and percentage of failed tasks for four different learning rates during the training of the agent. Since the learning rate is a crucial hyperparameter to train a neural network, we selected rates based on a range from 0.00001 to 0.001. Small learning rates cause a slow learning process, which is reflected in the case of 0.00001. Therefore, as shown in Figure \ref{scores}, the reward converges to its maximum value after episode 40 for this learning rate while other learning rates result in the same reward range after episode 20. On the other hand, increasing the learning rate does not affect the speed of the learning process in the case of 0.0001, 0.0005, and 0.001. The learning rate of 0.001 gives poor results after episode 60 because of the exploding gradient problem which is a result of both the linear activation function in the output layer and the neural network architecture including the number of layers and nodes. The exploding gradient problem causes extremely high weights in the neural network and therefore the model cannot learn anymore. It is also an indicator that the learning rate must be lowered. Hence, in our training phase, we focused on the other three learning rates considering 0.001 as the upper bound. Finally, we selected the final model which is trained by the learning rate of 0.0001 since it is the most successful regarding the total reward. The learning rates 0.0005 and 0.00001 are not as successful since they cannot exceed local optima due to their smaller values.

On the other hand, considering the rewards in the training phase, the upper bound is 75000 on average due to the total number of tasks for the scenario of 2400 mobile users. Similarly, the lower bound for rewards is -75000. Since we used the DDQN algorithm for the learning, the oscillation between episodes is smoother so that it is limited between rewards of 55000 and 45000. This is important since the time to converge would be higher in the traditional DQN algorithm as shown in \cite{van2016deep}. As a result, since the corresponding rewards do not significantly change, we choose our final model after the 100th episode.


\subsubsection{Competitors}

We evaluated the performance of DeepEdge considering our primary goal, which is minimizing the failed tasks. Hence, we applied four different approaches for the orchestration that are used in \cite{sonmez2019fuzzy} in order to examine the performance of the DeepEdge orchestrator appropriately. In these approaches, the \textit{network based} method first investigates the WAN bandwidth, which is initially 20 Mbps for each edge server. Afterwards, if the bandwidth is above the predefined threshold which is 6 Mbps, it forwards the task to the cloud. Otherwise, the task is offloaded to one of the edge servers that have the capacity to complete the task. On the other hand, the \textit{utilization based} approach checks the utilization of edge servers. If the total utilization of the edge servers is under 80\%, the task is offloaded to one of the edge servers in the network. However, if the total utilization of edge servers is higher than 80\%, the task is offloaded to the cloud. This threshold value assignment is based on a set of exploratory simulation experiments with different parameters. According to the simulation results, the best values for minimum WAN bandwidth and maximum CPU utilization are observed as 6 Mbps and 80\%, respectively. Apart from the results, the candidate threshold values were 12, 8, 6, 3 Mbps and 90, 80, 70, 60 percent for the minimum WAN bandwidth and the maximum CPU utilization, respectively. The \textit{hybrid} approach is a joint method of the network and utilization based techniques. Lastly, 
we used the \textit{fuzzy based} approach that is implemented in \cite{sonmez2019fuzzy} as the main competitor. We used the same input parameters in this study for the fuzzy logic process in order to perform a valid comparison.

\subsection{Experiments}

\subsubsection{Overall Results}

\begin{figure}[t]
\centering
\includegraphics[scale=0.63]{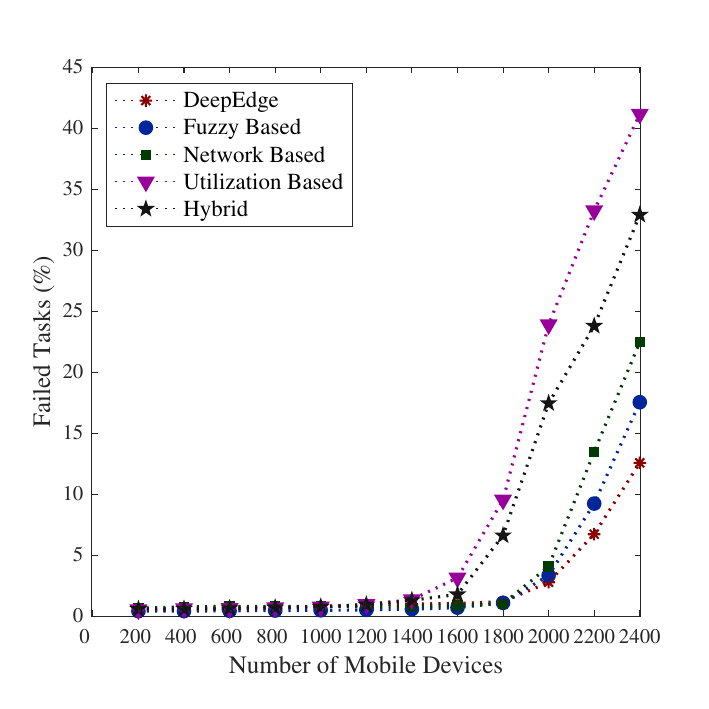}
\caption{The percentage of failed tasks}
\label{FailedTasksGeneral}
\end{figure}

We first examined the overall performance of DeepEdge in terms of the percentage of failed tasks. As shown in Figure \ref{FailedTasksGeneral}, the performance of all approaches is similar until 1600 mobile devices in the network. However, as the load builds up from 1600 to 2000 mobile devices, the accuracies of \textit{hybrid} and \textit{utilization based} approaches decrease. Especially when the number of mobile devices is higher than 1800, DeepEdge outperforms all other approaches. Considering the fact that the competitors are based on heuristic methods, and \textit{fuzzy logic} has 81 hand-crafted rules, it is a notable achievement that the agent explored the environment on its own and created the corresponding rules.

\begin{figure}[t]
\centering
\includegraphics[scale=0.63]{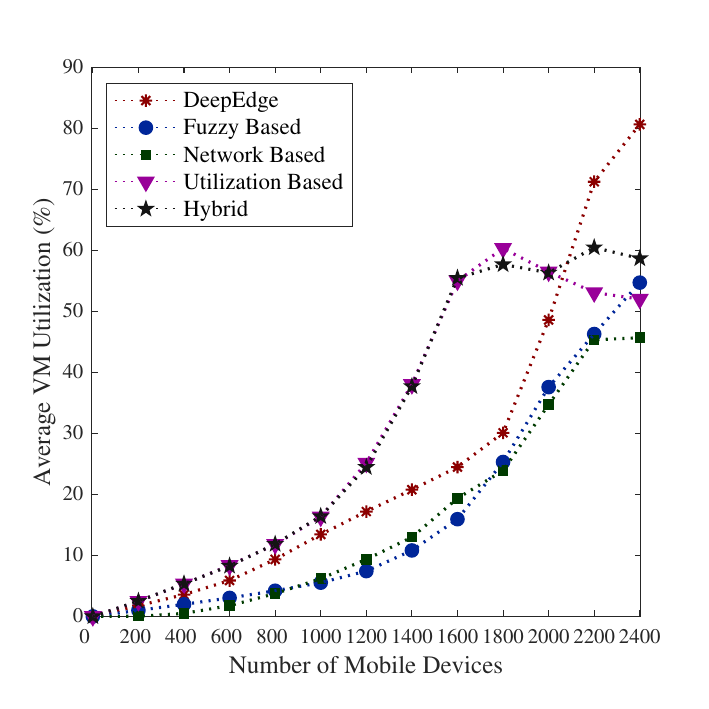}
\caption{Average VM utilization of edge and cloud servers based on the number of mobile devices.}
\label{VmUtilGeneral}
\end{figure}

We also analyzed how DeepEdge efficiently uses network and edge resources under high load. Figure \ref{VmUtilGeneral} shows that DeepEdge utilizes the resources in the network more efficiently than the other approaches, especially after 2000 users. This result is the effect of its accurate decisions for the offloaded tasks in the network considering the conditions of the edge servers and network bandwidth in long term. Since the decision of other approaches may cause congestion on the WAN or MAN, they cannot utilize computational resources well for the heavily-loaded cases. Thus, there is a ceiling effect in heavily-loaded cases since the overload on the network due to their inappropriate policies causes underutilized VMs on servers.

\begin{figure}[t]
\centering
\includegraphics[scale=0.63]{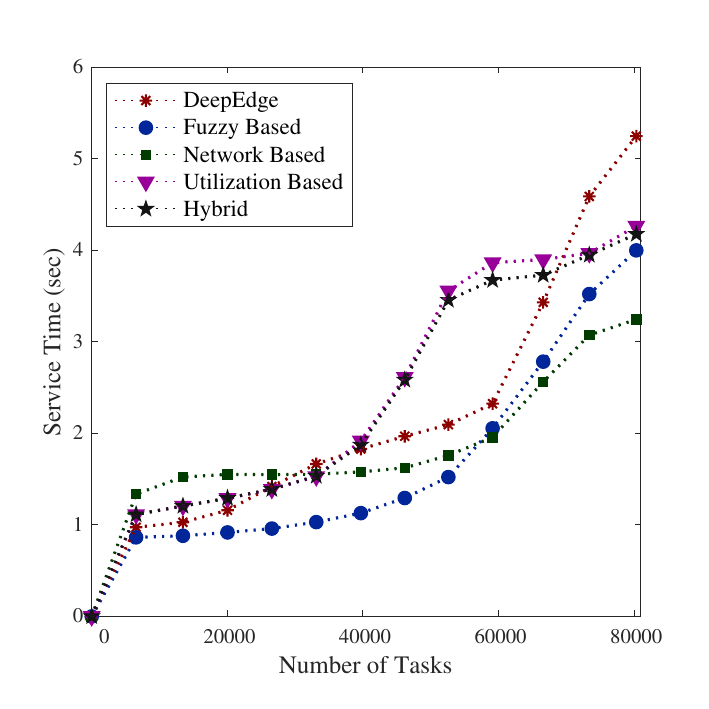}
\caption{The overall service time.}
\label{ServiceTime}
\end{figure}

\begin{figure}[t]
\centering
\includegraphics[scale=0.63]{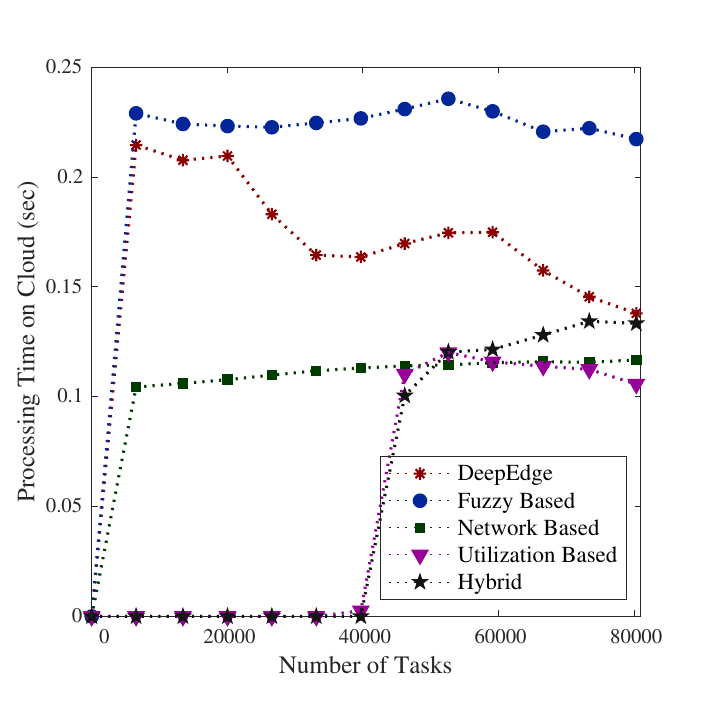}
\caption{The processing time on cloud.}
\label{ProcessingCloud}
\end{figure}

Next, we examined the overall service time for our proposed approach considering the network load in terms of tasks. Figure \ref{ServiceTime} exposes that the service time of DeepEdge is higher than the other approaches, particularly for the high number of tasks. Actually, this is not surprising since our objective function in Equation \ref{objectiveFunction} is that maximizing the success rate of the offloaded tasks, not minimizing the service time. Due to we design the reward mechanism based on our objective function, DeepEdge sacrifices the service time to maximize the cumulative reward. On the other hand, the cause of this effect can also be seen in Figure \ref{ProcessingCloud} more clearly in which we investigate the processing time on the cloud. It can be observed that DeepEdge offloads the tasks to the edge tier more when the number of tasks increase. This choice increases the overall service time in the end since some of the CPU-intensive apps' tasks would be sent to the edge rather than the cloud. However, it also provides the minimization of the failure rate, which is the goal of the DRL agent.

\subsubsection{Application-based Results}

\begin{figure*}[!t]

\begin{subfigure}{0.24\textwidth}
\includegraphics[width=\linewidth]{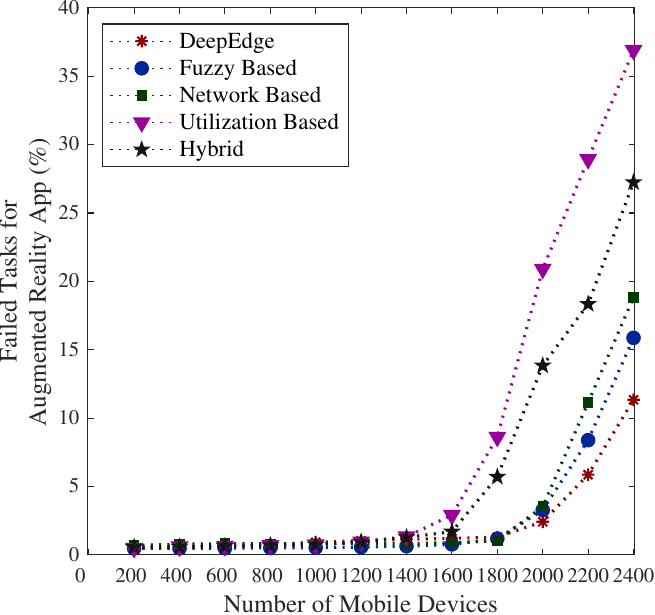}
\caption{Augmented Reality App} \label{AugmentedReality}
\end{subfigure}
\hspace*{\fill} 
\begin{subfigure}{0.24\textwidth}
\includegraphics[width=\linewidth]{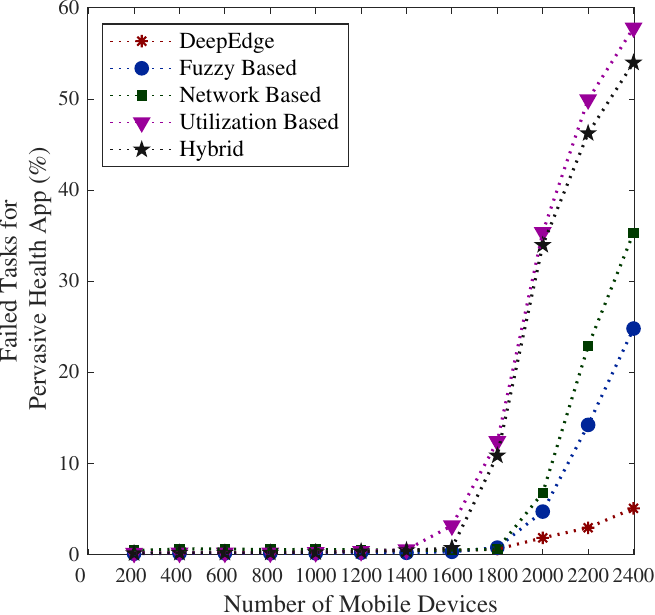}
\caption{Pervasive Health App} \label{HealthApp}
\end{subfigure}
\hspace*{\fill} 
\begin{subfigure}{0.24\textwidth}
\includegraphics[width=\linewidth]{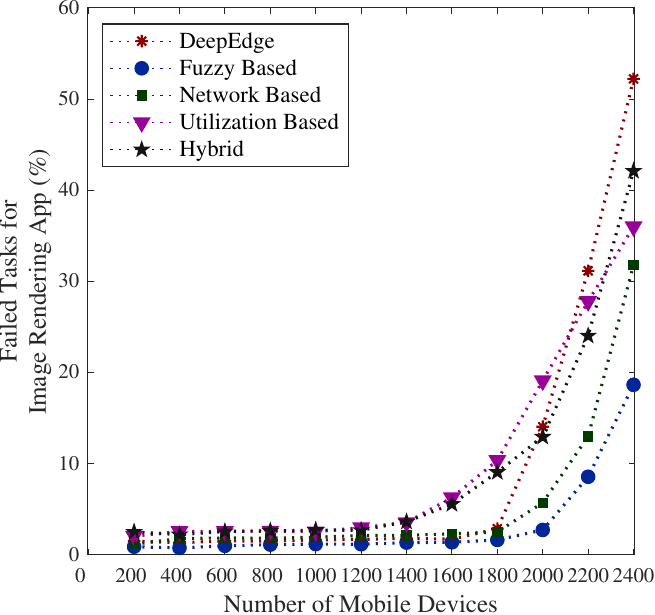}
\caption{Image Rendering App} \label{HeavyComp}
\end{subfigure}
\hspace*{\fill} 
\begin{subfigure}{0.24\textwidth}
\includegraphics[width=\linewidth]{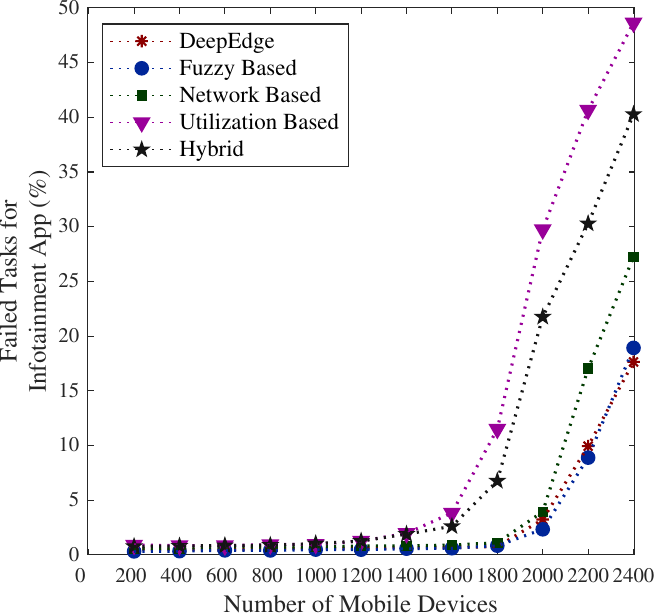}
\caption{Infotainment App } \label{InfotainmentApp}
\end{subfigure}
\caption{The change of failed tasks based on different application types. If the number of mobile devices increase in the network, DeepEdge outperforms other approaches in terms of successfully completed tasks. }
\label{FailedTasksApplication}
\end{figure*}

Based on the overall results, we then investigated the performance of DeepEdge for each application as shown in Figure \ref{FailedTasksApplication}. Considering the augmented reality, pervasive health, and infotainment applications, the robustness of DeepEdge manifests itself for the heavily-loaded environments. For the augmented reality application, our orchestrator offloads the generated tasks generally to the edge considering its delay sensitivity, which is relatively higher. The same requirement is also valid for the pervasive health application. However, since both applications cause different loads on a VM, our orchestrator also takes these properties into account regarding the \textit{TaskReqCapacity} feature. On the other hand, the delay sensitivity of the infotainment application is low. However, since its impact on VM utilization in terms of the load is high, the orchestrator must choose edge or cloud for the offloading wisely considering $t_{man}$ and $t_{wan}$. The results show that DeepEdge learned those requirements well. Thus, its performance is better than its competitors.

On the other hand, even though DeepEdge outperforms other approaches considering the different application types which have diverse task requirements, it cannot provide good results for the image rendering app. Therefore, in order to investigate this result, we performed several experiments using only the image rendering application. Hence, we trained three DDQN agents as the orchestrator by using the same hyperparameters given in Table \ref{Hyperparameters}. For these three agents, we only changed task interarrival times of the image rendering application for training. To this end, we applied 20 sec, as the original value, 10 sec, and 5 sec interarrival times. Finally, we compared the performance of DeepEdge with the Fuzzy Based approach as it gives the best performance for the image rendering application shown in Figure \ref{HeavyComp}.

\begin{figure*}[!t]

\begin{subfigure}{0.32\textwidth}
\includegraphics[width=\linewidth]{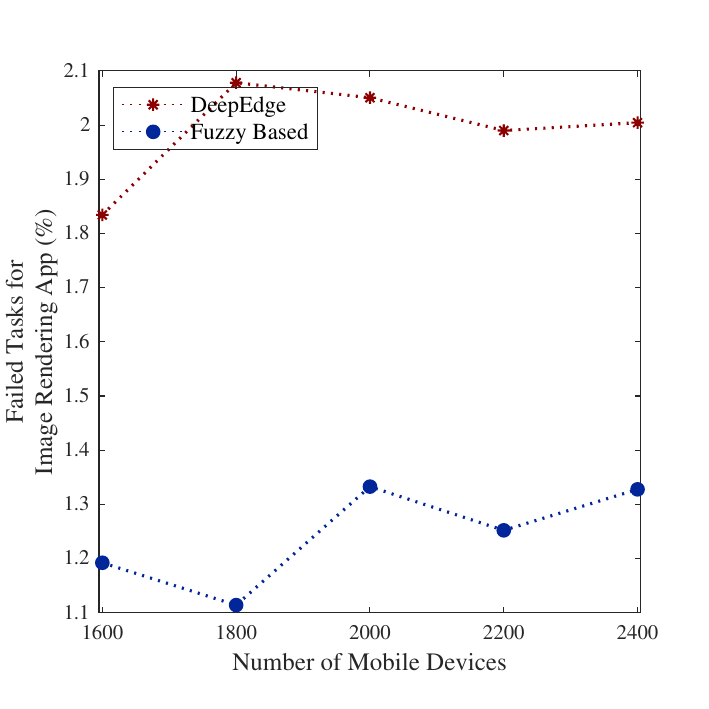}
\caption{20 sec task interarrival} \label{20sec}
\end{subfigure}
\hspace*{\fill} 
\begin{subfigure}{0.32\textwidth}
\includegraphics[width=\linewidth]{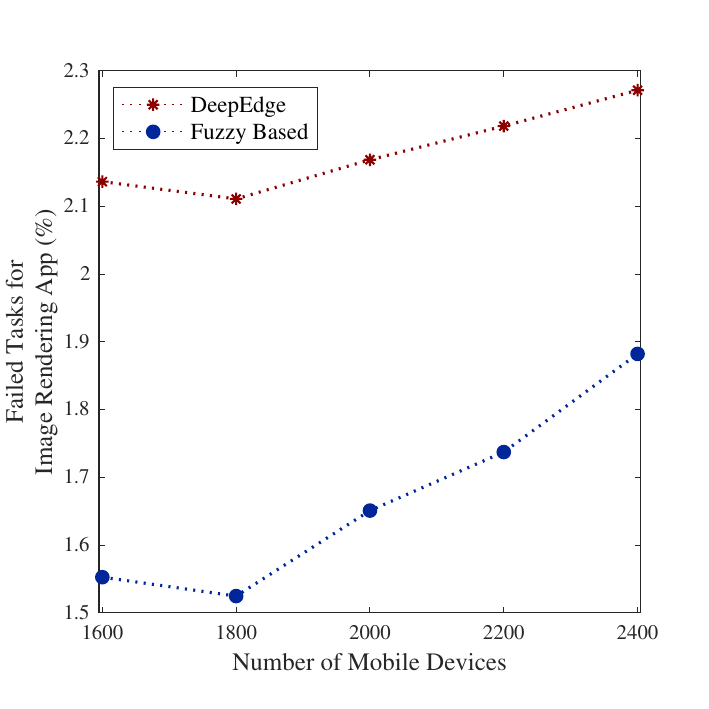}
\caption{10 sec task interarrival} \label{10sec}
\end{subfigure}
\hspace*{\fill} 
\begin{subfigure}{0.32\textwidth}
\includegraphics[width=\linewidth]{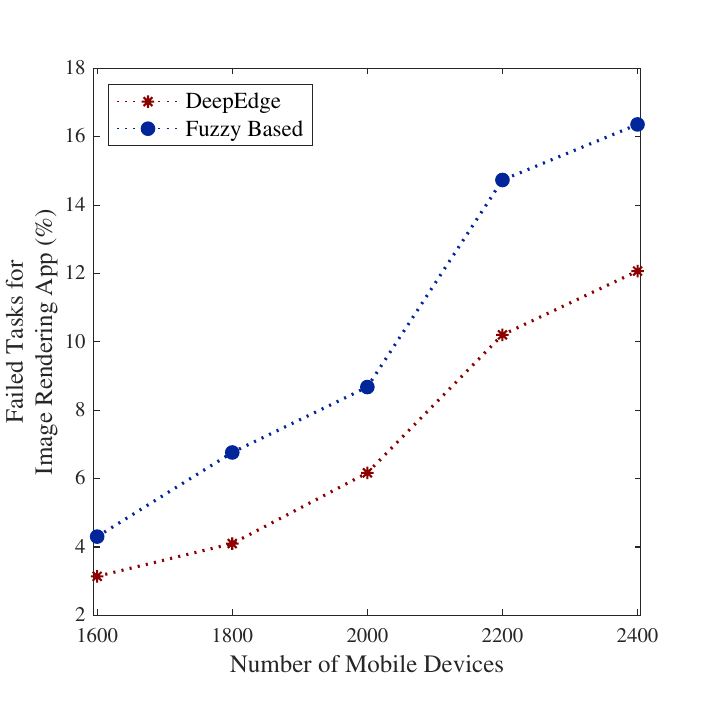}
\caption{5 sec task interarrival} \label{5sec}
\end{subfigure}
\caption{The effect of the task interarrival time on the image-rendering application.}
\label{FailedTasksImageRendering}
\end{figure*}

The results given in Figure \ref{FailedTasksImageRendering} show that DeepEdge outperforms the Fuzzy Based approach when the task interarrival time is 5 sec. On the other hand, there is no significant difference between the performance of 10 sec and 20 sec interarrival times. Thus, apart from the results in Figure \ref{FailedTasksImageRendering}, the reason that DeepEdge is not as successful for the image rendering app after 1800 users can be concluded by examining Table \ref{AppChars}. Since the average number of arriving image rendering tasks is 10, 6.66, and 2.85 times lower than the other applications respectively, states related to the decision of the image rendering tasks are relatively less for the training. Therefore, the agent could not learn the specific details of the requirements of the image rendering app. Furthermore, note that, since the image rendering tasks constitute only 3\% of the overall task offloading requests, the overall performance of DeepEdge is still significantly better for the heavily-loaded cases as shown in Figure \ref{FailedTasksGeneral}.

\section{Conclusion and Future Work}

We implemented a DRL based task orchestrator, namely DeepEdge, for edge computing. Throughout the study, we assumed that the offloading decision had been taken by mobile devices, and the task must be offloaded to one of the edge servers in the network or to the cloud. Due to the different characteristics of each application, generated tasks by mobile users must be handled by the orchestrator considering the application requirements and the current network conditions. For this reason, we developed DeepEdge by applying the DDQN algorithm. Even though providing MDP and coping with the delayed action effect are challenging due to the real-time, and stochastic nature of the edge computing environment, we successfully implemented a DRL agent based on DDQN for the orchestration. We conducted experiments in the EdgeCloudSim simulator by comparing our model with other approaches in the literature. Our results showed that DeepEdge has the capability to perform orchestration for different task types and outperforms its competitors. On the one hand, DeepEdge improved the average number of successfully completed tasks considering the different application requirements. On the other hand, DeepEdge used VM resources more efficiently since it had prevented the network bottlenecks so that offloaded tasks could arrive at their destinations. To the best of our knowledge, this is the first study in the domain of edge computing that evaluates the performance of a DRL model for different application types under various loads in terms of the number of mobile users.

For future work, we want to expand our study by applying federated learning in order to deal with the different task arrival rates of applications. To this end, we can perform the training in separate domains, and we can achieve better results for the application types whose task creation frequencies are relatively low in the network.


%



\section*{Acknowledgment}
This work is supported by the Turkish Directorate of Strategy and Budget under the TAM Project number 2007K12-873.


\ifCLASSOPTIONcaptionsoff
  \newpage
\fi



%

\bibliographystyle{IEEEtran}
\bibliography{DRL-Edge}

%

\begin{IEEEbiography}
[{\includegraphics[width=1in,height=1.25in,clip,keepaspectratio]{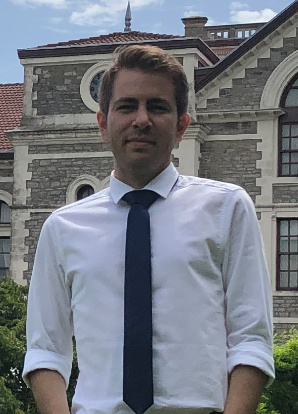}}]{Baris Yamansavacilar}
received his BS degree in Computer Engineering from Yildiz Technical University, Istanbul, in 2015. He received his MS degree in Computer Engineering from Bogazici University, Istanbul, in 2019. Currently, he is a PhD candidate and a research assistant in Computer Engineering Department at Bogazici University. His research interests include Edge Computing, Deep Reinforcement Learning, Mobile Networks, Software-Defined Networking, and Machine Learning.

\end{IEEEbiography}

\begin{IEEEbiography}[{\includegraphics[width=1in,height=1.25in,clip,keepaspectratio]{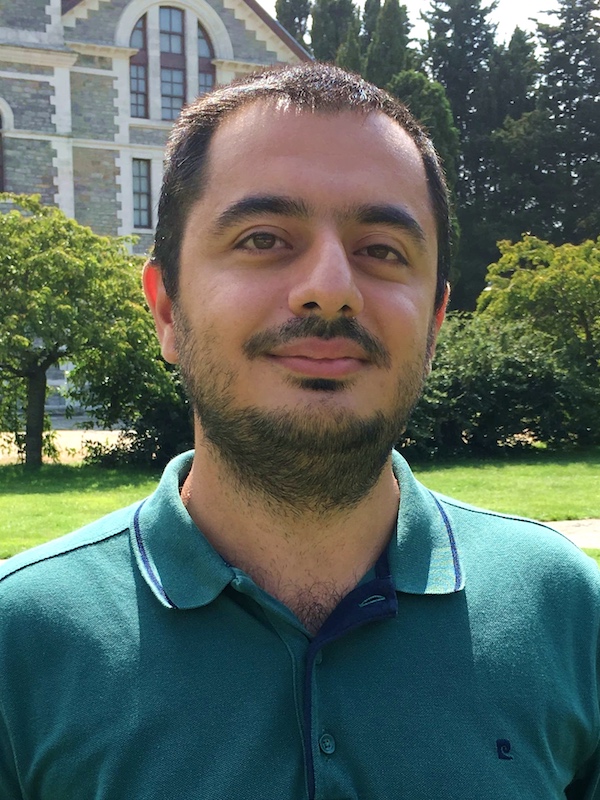}}]{Ahmet C. Baktir}
received his BS degree in Computer Science\&Engineering from Sabanci University, Istanbul, in 2012. He received his MA degree in Management Information Systems from Bogazici University, Istanbul, in 2014. He received a PhD in Computer Engineering in Bogazici University. He is a researcher in Ericsson Turkey. His research interests include Edge Computing, network management and orchestration.
\end{IEEEbiography}

\begin{IEEEbiography}
[{\includegraphics[width=1in,height=1.25in,clip,keepaspectratio]{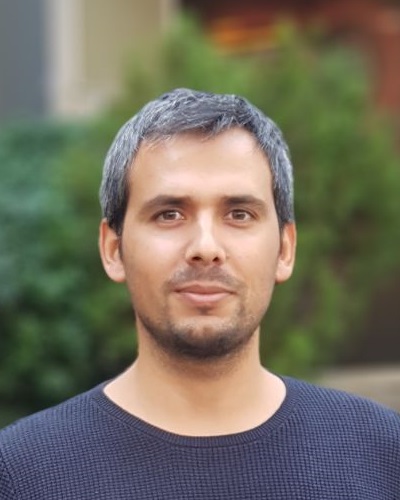}}]{Cagatay Sonmez}
received his BS degree in Computer Engineering from Dokuz Eylul University Izmir, in 2008. He received his MS and PhD degrees in Computer Engineering from Bogazici University, Istanbul, in 2012 and 2020. He has been working at Arcelik Electronics as a technical leader in R\&D Software Department, Istanbul since 2008. His research interests include design and performance evaluation of communication protocols, cloud computing, edge computing and IoT.
\end{IEEEbiography}

\begin{IEEEbiography}[{\includegraphics[width=1in,height=1.25in,clip,keepaspectratio]{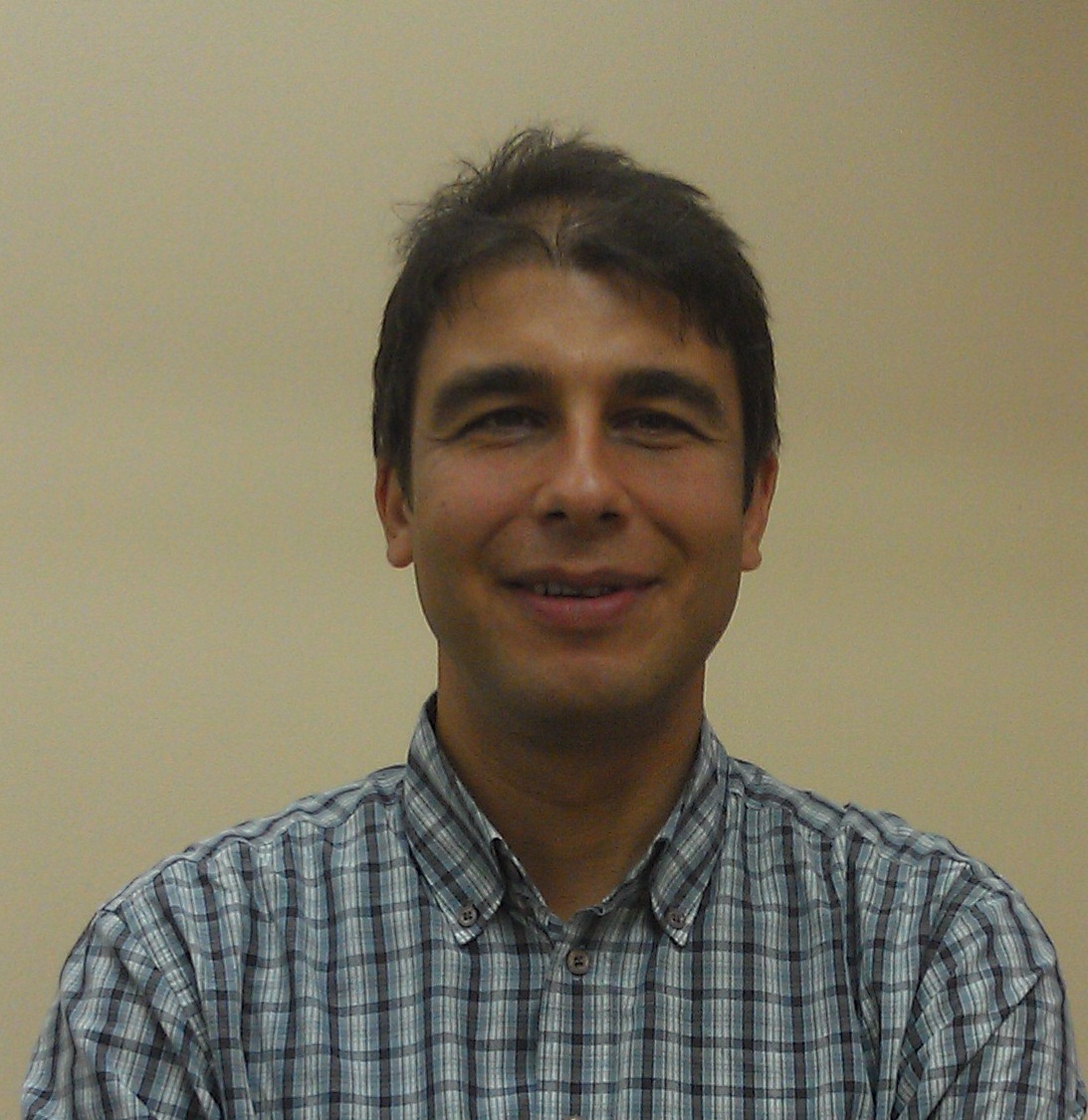}}]{Atay Ozgovde}
received his BS and MS degrees from Bogazici University, Istanbul, in 1995 and 1998, respectively. He worked as an R\&D engineer in NortelNetworks between 1998-2001. He completed his PhD degree in the NETLAB research group Bogazici University in 2009. He worked as a full-time faculty at the Galatasaray University Computer Engineering Department between 2009-2021. Currently, he is an assistant professor in the Computer Engineering Department, Bogazici University. His research interests include Computer Networks, Edge Computing, 5G and Beyond, Internet of Things, Ambient Intelligence, SDN and mobile cloud computing. He is a senior member of the IEEE.
\end{IEEEbiography}

\begin{IEEEbiography}[{\includegraphics[width=1in,height=1.25in,clip,keepaspectratio]{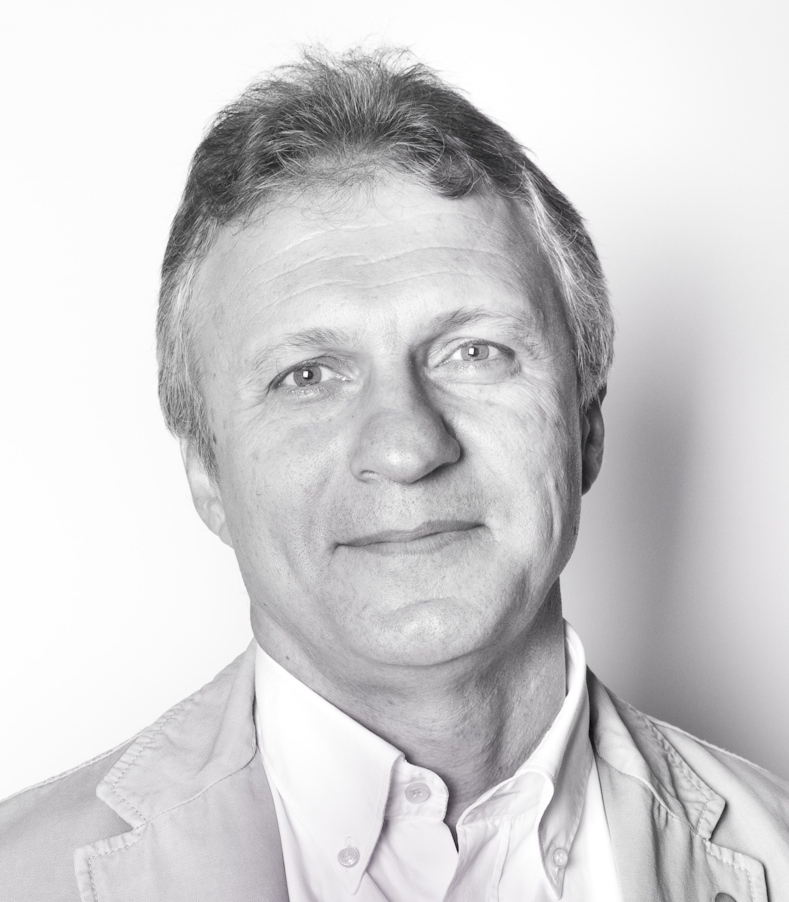}}]{Cem Ersoy}	
worked as an R\&D engineer in NETAS A.S. between 1984 and 1986. After receiving his PhD from Polytechnic University, New York in 1992, he became a professor of Computer Engineering and currently the department head in Bogazici University. Prof. Ersoy's research interests include wireless/cellular/adhoc/sensor networks, activity recognition and ambient intelligence for pervasive health applications, green 5G and beyond networks, mobile cloud/edge/fog computing, software defined networking, infrastructureless communications for disaster management. Prof. Ersoy is also the Vice Director of the Telecommunications and Informatics technologies Research Center, TETAM. Prof. Ersoy is a member of IFIP and was the chairman of the IEEE Communications Society Turkish Chapter for eight years.
\end{IEEEbiography}



\end{document}